\newcommand{\MuD}{\ensuremath{\mu{}\mathrm{D}}}

\newcommand{\FourHe}{\ensuremath{^{4}\mathrm{He}}}

\newcommand{\MuFourHe}{\ensuremath{(\mu{}^{4}\mathrm{He})^{+}}}
\newcommand{\MuThreeHe}{\ensuremath{(\mu{}^{3}\mathrm{He})^{+}}}

\newcommand{\OneSTwoS}{\ensuremath{\mathrm{1S\rightarrow{}2S}}}
\newcommand{\TwoSTwoP}{\ensuremath{\mathrm{2S\rightarrow{}2P}}}
\newcommand{\TwoS}{\ensuremath{\mathrm{2S}}}

\newcommand{\TwoP}{\ensuremath{\mathrm{2P}}}

\newcommand{\etal}{{\it et\,al.}}
\newcommand{\mev}{\,\ensuremath{\mathrm{meV}}}
\newcommand{\fm}{\,\ensuremath{\mathrm{fm}}}

\newcommand{\rh}{\ensuremath{r_h}}
\newcommand{\ra}{\ensuremath{r_\alpha}}

\def\OBACCA{a} 
\def\OFRIAR{b}
\def\OBORIE{c}
\def\OMART{d}
\def\ODATA{e}

\documentclass[aps,pra,twocolumn,superscriptaddress,nofootinbib,showkeys]{revtex4-1}

\usepackage{hyperref}
\usepackage[table,usenames,dvipsnames]{xcolor}
\usepackage{colortbl}
\usepackage[normalem]{ulem}
\usepackage{braket}
\usepackage{textcomp}
\usepackage{graphicx}
\usepackage{amsmath}
\usepackage{feynmp}
\usepackage{dcolumn}
\usepackage{multirow}

\definecolor{darkgreen}{rgb}{0,0.7,0.2}

\newcommand\REFEREE[1]{{#1}}

\newcommand\RP[1]{{#1}}

\newcommand\JI[1]{{#1}}

\newcommand{\iMPQ}{Max Planck Institute of Quantum Optics,  85748 Garching,
  Germany}
\newcommand{\iTRIUMF}{TRIUMF, 4004 Wesbrook Mall, Vancouver, BC V6T 2A3, Canada}
\newcommand{\iJGU}{Johannes Gutenberg-Universit\"at Mainz, QUANTUM, Institut f\"ur Physik \& Exzellenzcluster PRISMA, 55099 Mainz}
\newcommand{\iPSI}{Paul Scherrer Institute, 5232 Villigen-PSI, Switzerland}
\newcommand{\iETH}{Institute for Particle Physics and Astrophysics, ETH Zurich, 8093 Zurich,
  Switzerland}

\makeatletter
\newcommand\footnoteref[1]{\protected@xdef\@thefnmark{\ref{#1}}\@footnotemark}
\makeatother

\usepackage[para,online,flushleft]{threeparttable}

\newcolumntype{f}[1]{D{.}{.}{#1}}

\newcommand\cntr[2]{\multicolumn{#1}{c}{#2}}
\newcommand\lft[2]{\multicolumn{#1}{l}{#2}}

\newcommand{\app}{$\alpha$}
\newcommand{\Zap}{($Z\alpha$)}
\newcommand{\rs}{\ensuremath{\,r^{2}}}

\usepackage{lmodern}

\begin{document}

\pdfinfo {
/Title (Theory of the Lamb Shift and Fine Structure in Muonic Helium-4 Ions and the \textsuperscript{3}He--\textsuperscript{4}He Isotope Shift)
/Author (Julian J. Krauth)
}

\title{Theory of the Lamb Shift and Fine Structure in muonic \textsuperscript{4}He ions\\ and the muonic \textsuperscript{3}He--\textsuperscript{4}He Isotope Shift}

\author{Marc Diepold}
\affiliation{\iMPQ}

\author{Beatrice Franke}
\affiliation{\iMPQ}
\affiliation{\iTRIUMF}

\author{Julian J.~Krauth}
\email[Corresponding author: ]{jkrauth@uni-mainz.de}
\affiliation{\iMPQ}
\affiliation{\iJGU}

\author{Aldo Antognini}
\affiliation{\iETH}
\affiliation{\iPSI}

\author{Franz Kottmann}
\affiliation{\iETH}

\author{Randolf Pohl}
\affiliation{\iJGU}
\affiliation{\iMPQ}

\date{July 26, 2018}

\begin{abstract}

We provide an up to date summary of the \RP{theory} contributions to the \TwoSTwoP{} Lamb shift and the fine structure of the \TwoP{} state in the muonic helium ion \MuFourHe{}. 
This summary serves as the basis for the extraction of the alpha particle charge radius from the muonic helium Lamb shift measurements at the Paul Scherrer Institute, Switzerland.
Individual \RP{theory} contributions needed for a charge radius extraction are compared and compiled into a consistent summary.  The influence of the alpha particle charge distribution on the elastic two-photon exchange is studied to take into account possible model-dependencies of the energy levels on the electric form factor of the nucleus.
We also discuss the \RP{theory} uncertainty which enters the extraction of the \textsuperscript{3}He--\textsuperscript{4}He isotope shift from the muonic measurements. The \RP{theory} uncertainty of the extraction is much smaller than a present discrepancy between previous isotope shift measurements.
This work completes our series of $n=2$ theory compilations in light muonic atoms which we have performed already for muonic hydrogen, deuterium, and helium-3 ions.

\end{abstract}


\maketitle 

\section{Introduction}
\label{sec:intro}

\begin{figure}[b]
  \begin{center}
    \includegraphics[width=0.7\linewidth]{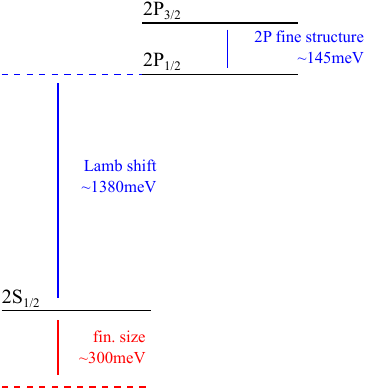}
  \end{center}
  \caption{The $\mathrm{2S}$ and $\mathrm{2P}$ energy levels in the muonic helium-4 ion. Since the nuclear spin is zero, no hyperfine structure is present. The figure is not to scale.}
  \label{fig:levels}
\end{figure}

The CREMA Collaboration has measured both \TwoSTwoP{} Lamb shift transitions in the muonic helium ion \MuFourHe{} \cite{Antognini:2011:Conf:PSAS2010,Nebel:2012:LEAP_muHe}. A scheme of these energy levels in muonic helium-4 ions is shown in Fig.\,\ref{fig:levels}. In preparation of the upcoming extraction of nuclear properties from these measurements, such as the nuclear root-mean-square (rms) charge radius \ra{},  
we provide a careful study of the available calculations of the theory contributions to the involved energy levels, summarizing the results of several theory groups.

Both, the Lamb shift and the fine structure, have been analyzed recently~\cite{Borie:2014:arxiv_v7,Krutov:2014:JETP120_73,Karshenboim:2012:PRA85_032509,Jentschura:2011:SemiAnalytic} but significant differences between the authors made it necessary to review the individual theory contributions. The same was previously done for muonic hydrogen~\cite{Antognini:2013:Annals}, muonic deuterium~\cite{Krauth:2016:mud}, and muonic helium-3 ions~\cite{Franke:2016:mu3HeTheo}.
Recent measurements of the \TwoSTwoP{} Lamb shift (LS) in other muonic atoms have already  provided the rms charge radii of the proton and the deuteron with unprecedented precision.
Results from muonic hydrogen measurements provided a proton charge radius of
\begin{equation}
  r_p^{\mu}~=~0.84087(26)^{\mathrm{exp}}(29)^{\mathrm{theo}}~\mathrm{fm}\ \text{\cite{Pohl:2010:Nature_mup1,Antognini:2013:Science_mup2}}. 
\end{equation}
This value is ten times more precise than the CODATA-2014 value of 0.8751(61)\,fm~\cite{Mohr:2016:CODATA14}, however also 4\,\%, or 6\,$\sigma,$ smaller.
This discrepancy created the so-called ``Proton-Radius-Puzzle'' (PRP)~\cite{Pohl:2013:ARNPS,Bernauer:2014:SciAm,Carlson:2015:Puzzle,Karshenboim:2015:mup,Krauth:2017:PRP}.
%

%

A recent determination of the deuteron radius from  muonic deuterium spectroscopy results in a value of
\RP{
\begin{equation}
\label{eq:Rd}
  r_d^{\mu}~=~2.12616(13)^{\mathrm{exp}}(89)^{\mathrm{theo}}~\mathrm{fm}\,\text{\cite{Pohl:2016:mud,Hernandez:2018:Rd_alive,Pachucki:2018:ThreePhoton}}, 
\end{equation}
}
that is also smaller than the CODATA value and hints towards a change in the Rydberg constant \cite{Pohl:2017:DSpec,Beyer:2017:2S4P},
\RP{but note the recent result~\cite{Fleurbaey:2018:1s3s}.}
\RP{The value in Eq.\,(\ref{eq:Rd}) differs slightly from our published value of 
$r_d^{\mu}~=~2.12562(13)^{\mathrm{exp}}(77)^{\mathrm{theo}}~\mathrm{fm}$\,\text{\cite{Pohl:2016:mud}}
due to updated nuclear theory of the 
two-photon contributions by 
Hernandez {\it et al.}~\cite{Hernandez:2018:Rd_alive}
and the unexpectedly large three-photon contribution
recently calculated for the first time by 
Pachucki {\it et al.}~\cite{Pachucki:2018:ThreePhoton}.
}
%
%

The determination of the alpha particle charge radius from muonic helium-4 ions, when compared to the radius determinations from electron scattering experiments \cite{Ottermann:1985:3He_4He_scatt,Sick:2008:rad_4He} will provide new input on the existing discrepancies. 
The improved value of $r_\alpha$ will be used in the near future for tests of fundamental bound state quantum electrodynamics (QED) by measurements of the \OneSTwoS{} transition in electronic He$^+$~ions\,\cite{Herrmann:2009:He1S2S,Kandula:2011:XUV_He}. 
Furthermore, the combination of the precise charge radii from  muonic helium-3 and helium-4 ions will contribute to \RP{solving} a discrepancy between several isotope shift measurements in electronic helium-3 and -4 atoms \cite{Shiner:1995:heliumSpec,CancioPastor:2012:PRL108,Rooij:2011:HeSpectroscopy,Zheng:2017:He_Iso}\footnote{The authors of Ref.\,\cite{Marsman:2015:QIshifts} point out that the values given by \cite{Shiner:1995:heliumSpec} and the experiment performed by \cite{CancioPastor:2012:PRL108}, might be affected by a systematic effect known as quantum interference~\cite{Horbatsch:2010:QI,Horbatsch:2011:PRA84,Marsman:2012:QIshifts,Brown:2013:QI,Marsman:2015:QIshifts,Amaro:2015:muonicQI,Amaro:2015:QIelliptical}.
Note that also the isotope shift value of \cite{Zheng:2017:He_Iso} is based on the measurement in \cite{CancioPastor:2012:PRL108}.
}.
And, finally, in combination with existing isotope shift measurements \cite{Wang:2004:6HeIso,Mueller:2007:HeIsotopes,Lu:2013:HeIso} the nuclear charge radii of the helium-6 and -8 isotopes will be slightly improved. 

A list of previous charge radius determinations is found in Angeli \etal{} \cite{Angeli:1999:radii} from 1999. A value from a combined analysis of experimental data is given in their more recent Ref.\,\cite{Angeli:2013:radii}. Their value of the alpha particle charge radius $r_\alpha = 1.6755(28)\fm$ is dominated by a measurement from Carboni \etal{} \cite{Carboni:1977:CollQuenchMup2S,Carboni:1978:LS_mu4he}, which has been excluded by a later measurement from Hauser \etal \cite{Hauser:1992:LS_search}. Hence, it should not be used. Instead the today established value is $r_{\alpha}=1.681(4)$\,fm, determined by Sick \cite{Sick:2008:rad_4He} from elastic electron scattering. \REFEREE{Note, that laser spectroscopy measurements of neutral helium atoms exist \cite{Shiner:1995:heliumSpec,Wang:2004:6HeIso,Mueller:2007:HeIsotopes,Rooij:2011:HeSpectroscopy,CancioPastor:2012:PRL108,Zheng:2017:He_Iso}, but 
\RP{as already mentioned above, these measurements yield only values}
for the isotope shift. For a precise absolute charge radius extraction from neutral helium atoms, theory calculations are not yet accurate enough~\cite{Patkos:2016:HeIso,Patkos:2017:HeIsoII}.}

The anticipated accuracy of the CREMA measurement will be about a factor of $\sim5$ more precise than the established value from electron scattering~\cite{Sick:2008:rad_4He,Sick:2015:JPCRD}.
The finite size effect in muonic helium, that is sensitive on the charge radius of the alpha particle, amounts to $\sim$300\,meV or 20\% of the LS in \MuFourHe{}.
The frequency uncertainty of the \MuFourHe{} LS measurements is on the order of 15\,GHz which corresponds to 0.06\,meV\footnote{1\,meV\,$\hat{=}$\,241.799\,GHz}.
In the following, this value serves as accuracy goal for the theory contributions. 
Several contributions have been calculated with uncertainties not much better than our accuracy goal, with the reasoning that the inelastic two-photon exchange (``polarizability'') will anyway dominate the extracted charge radius uncertainty. However, once a reliable value for the charge radius exists, e.g. from He or He$^+$ spectroscopy, these uncertainties will limit the extracted ``muonic polarizability''.
\RP{Further theory work is therefore warranted also for these ``pure QED'' 
terms,
because accurate values of the nuclear polarizability from muonic atoms may
eventually serve as important input for understanding the nuclear force
\cite{Ekstrom:2013:PRL110,Carlsson:2016:PRX,Hernandez:2018:Rd_alive}.} 
%

%
The paper is structured as follows:
Sec.\,\ref{sec:ls} discusses the pure QED contributions to the Lamb shift (independent of the charge radius and nuclear structure effects), which are summarized in Tab.\,\ref{tab:theory1}.
We use the theory calculations of Borie~\cite{Borie:2012:LS_revisited_AoP} (in this work we refer always to the updated Ref.\,\cite{Borie:2014:arxiv_v7}, version v7 on the arXiv) and the calculations of the group of Elekina, Faustov, Krutov, and Martynenko \etal~\cite{Krutov:2014:JETP120_73} (for simplicity referred to as ``Martynenko'' in the rest of the article) that provide a summary of terms contributing to the LS energy.
Various partial results of QED terms provided by the group of Ivanov, Karshenboim, Korzinin, and Shelyuto~\cite{Karshenboim:2012:PRA85_032509,Korzinin:2013:PRD88_125019} (for simplicity referred to as Karshenboim further on) and by Jentschura and Wundt~\cite{Jentschura:2011:SemiAnalytic} are compared. 
Sec.\,\ref{sec:ls2} discusses the contributions to the finite size effect, together with higher order corrections that scale with the nuclear charge radius squared $r_\alpha^2$. These charge radius dependent terms are summarized in Tab.\,\ref{tab:theory2}.
We use the works of Borie~\cite{Borie:2014:arxiv_v7}, Martynenko (Krutov~\etal\ \cite{Krutov:2014:JETP120_73}), and Karshenboim (Karshenboim~\etal\ \cite{Karshenboim:2012:PRA85_032509}).
Our summary provides the charge radius coefficient of the LS parameterization needed to extract the charge radius from the experimentally measured transitions.
In Sec.\,\ref{sec:tpe} we discuss the two-photon exchange (TPE) in \MuFourHe{}. 
%
In the first part, Sec.\,\ref{sec:friar}, we discuss the so-called nuclear and nucleon ``Friar moment'' contribution, also known as the third Zemach moment contribution.
%
%
%
%

%
In the second part, Sec.\,\ref{sec:pol}, we discuss the nuclear and nucleon polarizability contributions to the LS.
The nuclear polarizability, i.e.\ the inelastic part of the TPE stems from the virtual excitation of the nucleus and is related to its excitation spectrum~\cite{Holstein:1994:NuclPol}.
It is the least accurately known part of the \MuFourHe{} LS.
The TPE was recently investigated by the TRIUMF/Hebrew group in Ji \etal{} \cite{Ji:2013:PRL111,Ji:2018:nuclPOL}.
Recently also three-photon exchange contributions
have been discussed for hydrogen-like muonic atoms \cite{Pachucki:2018:ThreePhoton}, but yet no numbers for muonic helium exist.

The fine structure (FS) of the \TwoP{} state in \MuFourHe{} is studied in Sec.\,\ref{sec:fs} and summarized in Tab.\,\ref{tab:theory0}.
Our evaluation is based on the works of Borie~\cite{Borie:2014:arxiv_v7}, Martynenko (Elekina~\etal\ \cite{Elekina:2011:mu4He}), and Karshenboim (Karshenboim \etal\ \cite{Karshenboim:2012:PRA85_032509}, Korzinin \etal\ \cite{Korzinin:2013:PRD88_125019}).

In Sec.\,\ref{sec:isoShift}, we discuss the theory contributions with respect to the $^3$He--$^4$He isotope shift and extract a value for the uncertainty of the future value from the CREMA measurements in muonic helium ions. Here we exploit correlations between model-dependent calculations by the TRIUMF/Hebrew group to significantly reduce the theory uncertainty to the isotope shift. 
%
%

%
Throughout the paper we use the established convention and assign the \textit{measured} energy differences $\Delta E({\rm 2P_{1/2}-2S_{1/2}})$ and $\Delta E({\rm 2P_{3/2}-2S_{1/2}})$ a \textit{positive} sign.
%

%
Labeling of individual terms in LS and FS follows the convention of our
previous works~\cite{Antognini:2013:Annals,Krauth:2016:mud,Franke:2016:mu3HeTheo} 
in order to maintain comparability.
%
%
Terms that were found to not agree between various sources were averaged for our determination
and the resulting value is found in the ``Our Choice'' column.
These averaged values are given by the center of the covering band of all values 
under consideration $\nu{}_i$ with the uncertainty of their half spread, i.e.
\begin{equation}
  \begin{split}
    \mathrm{AVG}~~~=~~~ &\frac{1}{2}[\mathrm{MAX}(\nu{}_i)+\mathrm{MIN}(\nu{}_i)]\\
    \pm{}&\frac{1}{2}[\mathrm{MAX}(\nu{}_i)-\mathrm{MIN}(\nu{}_i)]
  \end{split}
\end{equation}
The values in the ``Our choice'' column are followed by the initial of the authors whose results were used to obtain this value (B\,=\,Borie, M\,=\,Martynenko \etal, K\,=\,Karshenboim \etal, J\,=\,Jentschura).
Important abbreviations: 
$Z$ is the nuclear charge, 
$\alpha$ is the fine structure constant.
``VP'', ``SE'', and ``RC'' refer to vacuum polarization, self energy, and recoil corrections, respectively. A proceeding $e$, $\mu$, or $h$ denotes contributions from electrons, muons, or hadrons.

\section{QED Lamb Shift in \MuFourHe{}}
\label{sec:ls}

\begin{figure}
  \begin{center}
    \input{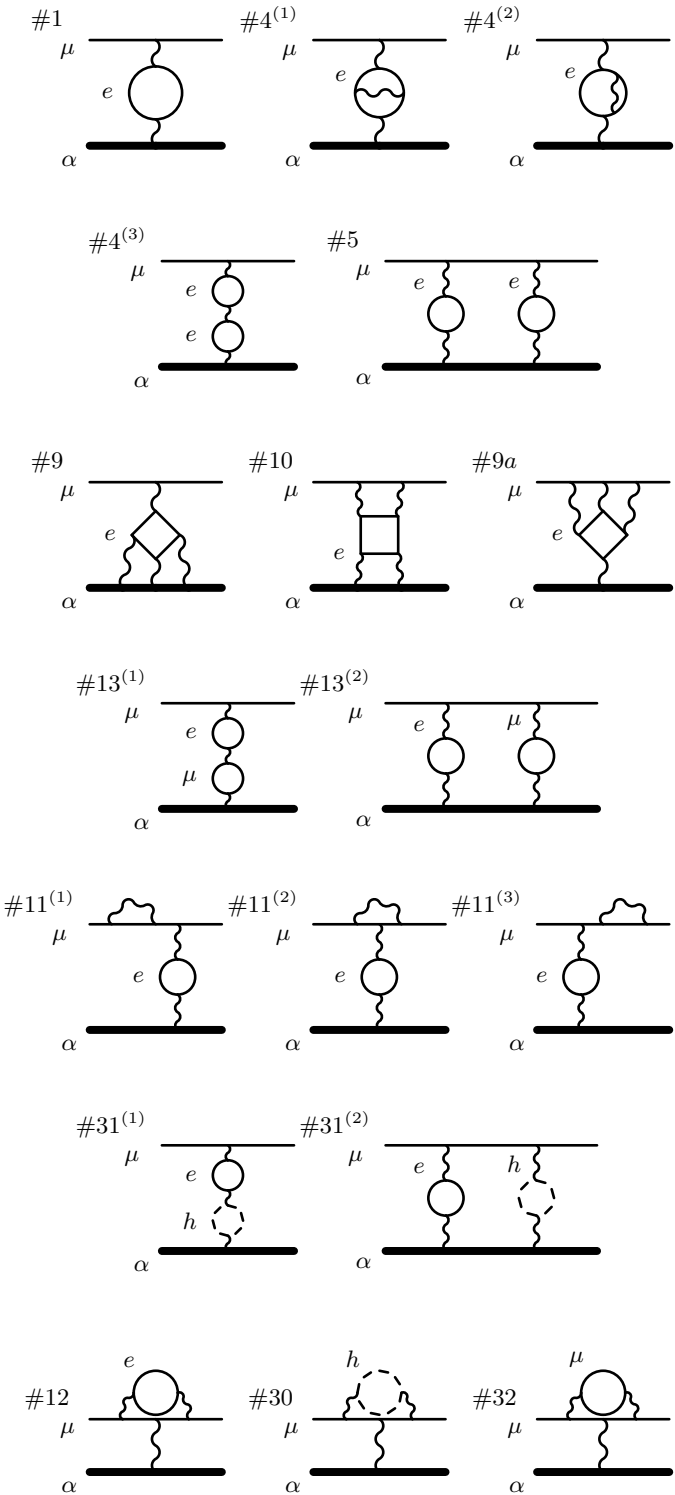}
    \caption{Important Feynman diagrams contributing to the QED part of the Lamb shift. \#1 The Uehling Term; \#4 The K\"all\'en-Sabry contribution; \#5 One loop eVP in two Coulomb lines; \#9/9a/10 Light-by-light scattering contributions; \#13 Mixed eVP/$\mu{}$VP; \#11 Self energy corr. to eVP; \#31 Mixed eVP/hadronic VP; \#12 eVP loop in SE contribution; \#30 Hadr. loop in SE contribution; \#32 $\mu$VP loop in SE contribution (included in \#21). In our summary, terms \#5, \#9, \#10, \#9a, \#13$^{(2)}$ and \#31$^{(2)}$ also contain their respective cross diagrams.}
    \label{fig:feynman}
  \end{center}
\end{figure}

First we consider pure QED terms that do not depend on nuclear properties.
All terms listed in this section are given in Tab.~\ref{tab:theory1}.
As in other muonic atoms, the one-loop electron vacuum polarization 
(eVP; \#1; see Fig.~\ref{fig:feynman})
is the largest term contributing to the Lamb shift.
Martynenko provides a non-relativistic calculation for this Uehling-term (\#1) together 
with a separate term for its relativistic corrections (Breit-Pauli correction, \#2)~\cite{Krutov:2014:JETP120_73}.
The values of the main term plus its correction (\#1$+$\#2) agree exactly with the independent
calculations of Karshenboim~\cite{Karshenboim:2012:PRA85_032509}
and Jentschura~\cite{Jentschura:2011:PRA84_012505,Jentschura:2011:SemiAnalytic} who follow the same procedure.
Borie's value (\#3) already includes relativistic corrections of
the order $\alpha{}(Z\alpha{})^2$ due to the use of relativistic Dirac wave functions~\cite{Borie:2014:arxiv_v7}.
The additional $\alpha{}(Z\alpha{})^4$ relativistic recoil correction to eVP (\#19) already included by the other authors
is treated separately in her framework~\cite{Borie:2014:arxiv_v7}.
The sum of all eVP contributions (\#1+\#2 or \#3+\#19) is in agreement 
between the calculations of all authors.
The average value of the Uehling contribution yields
\begin{equation}\label{eq:LS:leading}
  \Delta{}E_{\mathrm{(1-loop~eVP)}}= 1666.2946\pm0.0014\,\text{meV}.
\end{equation}
%
%
%
%

%
The next largest term in the QED part of the \MuFourHe{} Lamb shift is given by the two-loop electron vacuum polarization in the one-photon interaction of order $\alpha^2(Z\alpha)^4$ (\#4).
This so-called K\"all\'en-Sabry (KS) contribution is the sum of three Feynman diagrams
as seen in Fig.~\ref{fig:feynman}, \#4.
It is calculated by Borie and Martynenko, and the agreement between both calculations is 
still satisfactory, although not as good as for the Uehling term.
The one-loop eVP contribution with two Coulomb lines 
(\#5, see Fig.~\ref{fig:feynman}) is calculated by Martynenko~\cite{Krutov:2014:JETP120_73} 
and Jentschura~\cite{Jentschura:2011:SemiAnalytic} and their results show
satisfactory agreement.
Borie cites a paper of Karshenboim\,\cite{Karshenboim:2012:PRA85_032509}, it is however unknown how she obtained the quoted value.
Karshenboim calculates the sum of both terms (\#4+\#5)~\cite{Korzinin:2013:PRD88_125019}.
The sum is in excellent agreement with the sum of Martynenko's values and also agrees with the calculation of Borie.
The total contribution from two eVP loops in one and two Coulomb lines is given by the average of 
\RP{
\begin{equation}
  \label{eq:avg:45}
  \Delta{}E_{\mathrm{(2-loop~eVP~1\&2~C-lines)}}= 13.2794\pm0.0026\,\text{meV}.
\end{equation}
}

%
%

Calculations of third order eVP contributions (3-loop eVP, \#6+\#7, from Martynenko and Karshenboim, agree for the required accuracy~\cite{Borie:2014:arxiv_v7,Krutov:2014:JETP120_73,Korzinin:2013:PRD88_125019}. 
Karshenboim's value is chosen because he showed that the calculation method of Martynenko, first employed by Kinoshita and Nio\,\cite{Kinoshita:1999:ThreeLoop}, is not correct\,\cite{Korzinin:2013:PRD88_125019}.
The value of the third order eVP is given by 
\RP{
\begin{equation}
  \label{eq:avg:67}
  \Delta{}E_{\mathrm{(3-loop~eVP)}}= 0.0740\pm0.0030\,\text{meV}.
\end{equation}
}
%
%
The size of the third order contribution is comparable in size to our
accuracy goal while the uncertainty of the term is even smaller.  
The contribution from two eVP loops in one and two Coulomb lines has additional RC 
of the order $\alpha{}^2(Z\alpha{})^4m$ (\#29).
This correction was calculated by Martynenko and Karshenboim, but the calculations 
differ by more than a factor of two~\cite{Krutov:2014:JETP120_73,Korzinin:2013:PRD88_125019}.
Since similar calculations for \MuD{} are in agreement in previous 
publications of both authors~\cite{Martynenko:2014:muD_Theory,Korzinin:2013:PRD88_125019} further investigation
is needed.
For our summary we choose the average value of $0.0039\pm0.0018$\,meV due to the 
disagreement between Martynenko and Karshenboim.
Fortunately, the small discrepancy is not relevant on the level of the accuracy goal.
%

%
The higher order ``light-by-light'' scattering contribution 
consists of three individual terms (\#9,\#10,\#9a; see Fig.\,\ref{fig:feynman}).
The first, so-called Wichmann-Kroll term, (\#9) is in agreement between 
the works of Karshenboim~\cite{Karshenboim:2010:JETP_LBL} and Martynenko~\cite{Krutov:2014:JETP120_73}.
Borie~\cite{Borie:2014:arxiv_v7} also calculates the term independently and reports a slightly smaller but still agreeing result.
The remaining Virtual Delbr\"uck (\#10) and inverted Wichmann-Kroll (\#9a)
contributions have been calculated by Karshenboim~\cite{Karshenboim:2010:JETP_LBL}.
The Virtual Delbr\"uck contribution was calculated by Borie earlier\,\cite{Borie:1976:Virtual} although with larger uncertainty. As can be seen from Tab.~\ref{tab:theory1}, cancellations between the three terms occur. We therefore directly adopt Karshenboim's values to include all cancellations.
A term comparable in size to the K\"all\'en-Sabry contribution (but with opposite sign) is given by the effect of muon vacuum polarization ($\mu$VP) and muon self energy ($\mu$SE) (\#20).
The sum of both terms was calculated by Borie and Martynenko and they show satisfactory agreement~\cite{Borie:2014:arxiv_v7,Krutov:2014:JETP120_73}.
$\mu$SE also contributes as correction to the one-loop eVP term (see Fig~\ref{fig:feynman}, \#11). The results of Karshenboim and Jentschura agree very well~\cite{Karshenboim:2010:JETP_LBL,Jentschura:2011:SemiAnalytic}. The calculation from Martynenko in Ref.~\cite{Krutov:2014:JETP120_73} provides an incomplete value since he only calculates the second diagram seen in Fig.~\ref{fig:feynman}, \#11$^{(2)}$. He adopts the complete value of Jentschura in his summary. Borie only partially calculates this term, as stated in appendix C of her summary~\cite{Borie:2014:arxiv_v7}. Therefore our choice is compiled from Karshenboim's and Jentschura's value.
Insertion of an eVP or hVP loop in the $\mu$SE correction leads to corrections of higher order. The contribution of the additional eVP loop (\#12) was calculated by Borie and Karshenboim and their values agree well. The hVP term (\#30) was only calculated by Karshenboim whose value we adopt. 
There is also a contribution due to a $\mu$VP insertion in the $\mu$SE line. This contribution is not separately added to our summary, because it is already included in the $\mu$SE value. 
The contribution with an eVP and a $\mu$VP loop in the one photon interaction is given by the first diagram of \#13 in Fig.~\ref{fig:feynman}. It was evaluated by Martynenko and Borie and their values agree.
Karshenboim provides values of this contribution summed with the respective term in the two Coulomb line diagram (second part of \#13)~\cite{Karshenboim:2010:JETP_LBL}.
Both terms are of similar size, therefore the values of Karshenboim and
Borie/Martynenko differ by nearly a factor of two.
Since the total term is small, this uncertainty is not important for the
Lamb shift extraction.
In addition, Karshenboim also calculated the influence of the mixed eVP-hVP
diagram in one and two Coulomb lines (Fig.\ref{fig:feynman}, \#31).
Borie only gives a term labeled ``higher order correction to $\mu$SE and $\mu$VP'' 
(\#21) that also includes the $\mu$VP loop in the SE contribution (previously \#32).

%
%
%
%

%
The insertion of a hadronic vacuum polarization (hVP; \#14) loop in the one Coulomb-photon 
interaction leads to another correction calculated by Borie and Martynenko.
Both values agree within the uncertainty given in Borie's publication~\cite{Borie:2014:arxiv_v7}.
We use Borie's result~\cite{Borie:2014:arxiv_v7} as her uncertainty includes Martynenko's value\,\cite{Krutov:2014:JETP120_73}.
Item \#17 is the main recoil correction in the Lamb shift, also called
the Barker-Glover correction.
%
%
The available calculations of the term by Borie, Martynenko and Karshenboim agree perfectly.

Item (\#18) is the term called ``recoil finite size'' by Borie~\cite{Borie:2014:arxiv_v7}. It is of order $(Z\alpha)^5\braket{r}_{(2)}/M$ and is linear in the \textit{first} Zemach moment. It has first been calculated by Friar ~\cite{Friar:1978:Annals} (see Eq.\,F5 in App.\,F) for hydrogen and has later been given by Borie \cite{Borie:2014:arxiv_v7} for $\mu$d, \MuThreeHe, and \MuFourHe.
We discard item \#18 because it is considered to be included in the elastic TPE \cite{Pachucki:PC:2015,Yerokhin:2016:RCFS}.
%
%

%
Further relativistic recoil corrections of the order $(Z\alpha{})^5$ and $(Z\alpha{})^6$ are also
included in our summary (\#22, \#23).
The $(Z\alpha{})^5$ correction was calculated by Borie, Martynenko and Jentschura and their results agree.
The $(Z\alpha{})^6$ term was only determined by Martynenko, but is two orders of magnitude smaller than
the term of the previous order. Therefore we simply accept his value in our summary.
Martynenko provides a term called "radiative correction with recoil of the order
$\alpha(Z\alpha{})^5$ and $(Z^2\alpha{})(Z\alpha{})^4$".
The respective terms are included in Borie's ``higher order recoil'' together
with some additional terms not covered by Martynenko.
We therefore adopt the more complete value of Borie (\#24).
The total logarithmic recoil in \MuFourHe{} of the order $\alpha(Z\alpha{})^5$
(\#28)  was only calculated by Jentschura~\cite{Jentschura:2011:SemiAnalytic}.
It includes the dominant seagull-term as well as two more Feynman-diagrams
with smaller contributions. 
We directly adopt this result for our summary.
From the summary given in Tab.~\ref{tab:theory1} we extract the total nuclear 
structure independent part of the Lamb shift
\begin{equation}\label{eq:LS:tot}
  \Delta{}E_{(LS,\,QED)}= 1668.4892\pm0.0135\mev.
\end{equation}
This value is in agreement with the sum given by Martynenko~\cite{Krutov:2014:JETP120_73},
and agrees also with Borie's value when discarding the recoil finite size term.
In the case of \MuFourHe{} there is no Darwin-Foldy (DF) term as opposed to
$\mu$D~\cite{Krauth:2016:mud}.
This term normally accounts for the Zitterbewegung of the nucleus but
vanishes in \MuFourHe{} due to its zero nuclear spin.
The uncertainty of the ``pure QED'' contributions in Eq.\,(\ref{eq:LS:tot}) is a factor of 4 smaller than the expected experimental uncertainty and poses no limitation of the charge radius extraction.

\section{{r}$^2$ contributions to the Lamb Shift}
\label{sec:ls2}

The \TwoSTwoP{} splitting is also affected by the charge radius of the alpha particle. This so-called finite size effect dominantly influences S states due to their non-zero wave function at the origin, $\Psi{}(0)$. The finite size contributions in \MuFourHe{} can be parameterized with the square of the nuclear root-mean-square (rms) charge radius, which is defined as \cite{Jentschura:2011:DF,Pohl:2013:ARNPS}
\begin{equation}
\ra^2 = -6\frac{dG_{E}}{dQ^2}\big|_{Q^2=0}\,
\end{equation}
where $G_{E}$ is the Sachs electric form factor of the nucleus and $Q^2$ is the square of the four-momentum transfer to the nucleus. This charge radius definition is consistent with the one used in elastic electron scattering. In a simplified, nonrelativistic picture the nuclear charge radius is often referred to as the second moment of the nuclear charge distribution.

The leading order finite size contribution (\#r1) is of order $(Z\alpha{})^4$ and originates from the one-photon interaction between the muon and the helium nucleus. It is calculated by inserting the form factor in the nucleus vertex. 
The coefficient of the leading order finite size effect is provided by Borie\,\cite{Borie:2014:arxiv_v7} and Karshenboim\,\cite{Karshenboim:2012:PRA85_032509}, and their results agree.
Martynenko\,\cite{Krutov:2014:JETP120_73} however only gives absolute energy values for the finite size effect. For the leading order contribution he obtains $-295.85\pm2.83\mev$.
We have to divide by the square of the rms charge radius of 1.676(8)\,fm used in his calculations \cite{Krutov:2014:JETP120_73}, to get the resulting coefficient given in Tab.~\ref{tab:theory2}. Martynenko's value agrees with the other two.
All authors follow the previous calculations of Friar\,\cite{Friar:1978:Annals}.
\#r1 is given by the average of the three authors as

\begin{equation}\label{eq:finsize:r1}
  \Delta E(\mathrm{\#r1})= -105.3210\pm0.0020\,\mev/\fm^2\,\ra^2,
\end{equation}
where the uncertainty is far better than our accuracy goal.

%
Item \#r4 is the one-loop $e$VP (\textit{Uehling}) correction of order $\alpha(Z\alpha)^4$, i.e.\ an $e$VP insertion into the one-photon line. It has been calculated by all three groups, Borie\,\cite{Borie:2014:arxiv_v7}, Martynenko\,\cite{Krutov:2014:JETP120_73} and Karshenboim\,\cite{Karshenboim:2012:PRA85_032509}. On p.\,31 of \cite{Borie:2014:arxiv_v7}, Borie notes that she included the correction arising from the K\"all\'en-Sabry (KS) potential in her $b_d$. This means that her value already contains item \#r6, which is the two-loop $e$VP correction of order $\alpha^2(Z\alpha)^4$. Item \#r6 is given explicitly only by the Martynenko group \cite{Krutov:2014:JETP120_73} (No.\,18, Eq.\,73). The sum of Martynenko \etal's \#r4 and \#r6 differs by 0.014\mev/fm$^2$ from Borie's result. Using a charge radius of 1.681\,fm this corresponds to roughly 0.04\mev{} and, hence, causes the largest uncertainty in the radius-dependent one-photon exchange part. The origin of this difference is not clear \cite{Borie:PC:2017,Martynenko:PC:2017}. A clarification of this difference is desired but does not yet limit the extraction of the charge radius. As \textit{our choice} we take the average of the sum (\#r4+\#r6) of these two groups. The resulting average does also reflect the value for \#r4 provided by Karshenboim \etal~\cite{Karshenboim:2012:PRA85_032509}.

Item \#r5 is the one-loop $e$VP (\textit{Uehling}) correction in second order perturbation theory (SOPT) of order $\alpha(Z\alpha)^4$. It has been calculated by all three groups,
Borie\,\cite{Borie:2014:arxiv_v7}, Martynenko\,\cite{Krutov:2014:JETP120_73} and Karshenboim\,\cite{Karshenboim:2012:PRA85_032509}. On p.\,31 of \cite{Borie:2014:arxiv_v7}, Borie notes that she included the two-loop corrections to $\epsilon_{VP2}$ in her $b_e$. This means that her value already contains item \#r7, which is the two-loop $e$VP in SOPT of order $\alpha^2(Z\alpha)^4$. 
Item \#r7 is only given explicitly by the Martynenko group \cite{Krutov:2014:JETP120_73} (No.\,19). The sum of Martynenko \etal's \#r5+\#r7 differs by 0.01\mev{} from Borie's result. As \textit{our choice} we take the average of the sum (\#r5+\#r7) of these two groups.
Again here, {\it our choice} reflects the value for \#r5 provided by Karshenboim \etal~\cite{Karshenboim:2012:PRA85_032509}.

The nuclear structure influence on the 2P$_{1/2}$ state is only determined by Borie (\#r8)\,\cite{Borie:2014:arxiv_v7}.
We directly adopt her value, but change the sign of \#r8 from the original publication to be consistent with our nomenclature of tabulating $\Delta{}E(\mathrm{2P-2S})$.

Item \#r2 is a radiative correction of order $\alpha(Z\alpha{})^5$. It has been calculated by Borie\,\cite{Borie:2014:arxiv_v7} and Martynenko\,\cite{Krutov:2014:JETP120_73}. Their values agree well.
Martynenko\,\cite{Faustov:2017:rad_fin_size} recently calculated an additional term which we denote as \#r2'. It has a non-trivial dependence of the charge radius and is therefore provided as an absolute value rather than as a coefficient. Since it is tiny this procedure does not affect the extracted charge radius. Note, that in \cite{Faustov:2017:rad_fin_size}, Martynenko indicates the value for the 1S state, which has to be scaled by $1/2^3$ to account for the 2S state.

%
Item \#r2b' is a VP correction of order $\alpha(Z\alpha)^5$. It is an elastic contribution and only calculated by the Martynenko group \cite{Krutov:2014:JETP120_73}. It is not parameterized with the charge radius squared and therefore given as a constant. We do \textit{not} include this correction for the following reason: In muonic deuterium this correction cancels to a large amount with its inelastic counterpart \cite{Krutov:2011:PRA84_052514} (see below Eq.\,(62)). This cancellation is also expected for muonic helium ions. Since the inelastic correction of same order has not been calculated yet, we decided to not include this value in the final sum.

For the finite size term of the order  $(Z\alpha{})^6$ (\#r3) and the same-order correction (\#r3'), Borie and Martynenko use different methods of calculation. Here, \#r3' is given as an absolute value, because of its non-trivial dependence on the charge radius, similar to \#r2'.
A term corresponding to the $\braket{\mathrm{ln}\,r}$ coefficient is part of term (\#r3) for Martynenko and part of (\#r3') for Borie, leading to a correlation between both.
In order to stay consistent with the summary in \MuD{}\,\cite{Krauth:2016:mud} we decided to average both terms providing $\#r3=-0.1340\pm0.0030$\,meV/fm$^2$ and $\#r3'= 0.067\pm0.012$\,meV until a clear definition is settled on.
Note that although the uncertainty of \#r3' is still a factor of 5 smaller than the uncertainty goal, it would be helpful if this 20\% relative uncertainty in the term could be improved.

The total $\ra^2$ coefficient of the Lamb shift is given by
\begin{equation}\label{eq:finsize:tot}
\begin{split}
  \Delta{}E_{(Fin.\,size)}= &-106.3536(82)\mev/\fm^2\,\ra^2\\
  &+ 0.0784(112)\mev.
\end{split}
\end{equation}
The uncertainty of the first term corresponds to 0.02\mev\ (for $\ra=1.681\fm$), already 30\% of our uncertainty goal.
%
%
%

\section{Two-photon exchange}
\label{sec:tpe}

\begin{center}
  \begin{figure}
    \input{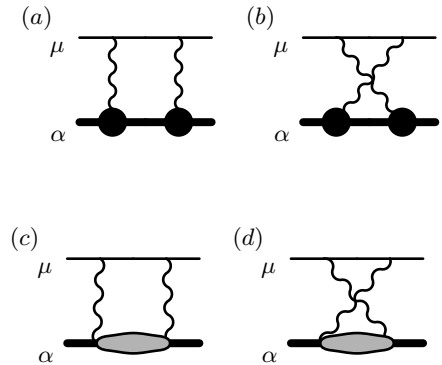}
    \vspace{10mm}
    \caption{Two-photon exchange in the \MuFourHe{} Lamb shift.
    The two diagrams $(a)+(b)$ are contributing to the Friar moment contribution 
    $\delta{}E_{\mathrm{Friar}}$ of the Lamb shift while 
    the similar diagrams $(c)+(d)$ show the nuclear polarizability contribution of
    the helium nucleus $\delta{}E_{\mathrm{inelastic}}$. Thick dots
    indicate form factor insertions while the gray blobs represent all
    possible excitations of the nucleus.}
    \label{fig:feynman2}
  \end{figure}
\end{center}
Important parts of the nuclear structure dependent Lamb shift contributions
are created by the two-photon exchange (TPE) between muon and nucleus (see Fig.~\ref{fig:feynman2}). 
Two distinct parts can be separated:
\begin{equation}
  \Delta{}E_{\mathrm{TPE}}^{\mathrm{LS}}=\delta{}E_{\mathrm{Friar}}^{A+N} + \delta{}E_{\mathrm{inelastic}}^{A+N},
  \label{eq:TPE}
\end{equation}
where $\delta{}E_{\mathrm{Friar}}^{A+N}$ is the Friar moment contribution\,\footnote{The term ``Friar moment'' has been introduced by Karshenboim \etal\ in \cite{Karshenboim:2015:PRD91_073003}.} (Fig.~\ref{fig:feynman2}\,(a)+(b)), also known as ``third Zemach moment contribution'', and 
$\delta{}E_{\mathrm{inelastic}}^{A+N}$ is the inelastic part of the TPE, also called the polarizability contribution (Fig.~\ref{fig:feynman2}\,(c)+(d)). Each part can again be separated into a nuclear (\textit{A}) and a nucleon (\textit{N}) part.

\subsection{The Friar moment contribution in \MuFourHe{}}
\label{sec:friar}

The Friar moment contribution $\delta{}E_{\rm Friar}^{A}$ is an elastic contribution, analog to the finite size effect, but of order $(Z\alpha)^5$, i.e.\ in the two-photon interaction (see Fig.\,\ref{fig:feynman2}, $(a),(b)$).
In the following we discuss five ways of how the Friar moment can be obtained:

\JI{
\textit{Option \OBACCA}: The most modern calculation of the Friar moment contribution is provided by the TRIUMF/Hebrew group \cite{Ji:2014:FewBodySyst,Hernandez:2016:POLupdate,Ji:2018:nuclPOL}. They obtain the nuclear Friar moment contribution $\delta E_{\rm Friar}^A$ by performing \textit{ab initio} calculations, using state-of-the-art nuclear potentials.
Their  result of \cite{Ji:2018:nuclPOL}
\begin{equation}\label{eq:obacca_nuclear}
  \delta E_{\rm Friar}^A(\OBACCA) = 6.14 \pm 0.31\,\mev
\end{equation}
uses the sum of their terms $\delta{}_{Z1}$ and $\delta{}_{Z3}$ \cite{Ji:2014:FewBodySyst} as an approximation for the elastic Friar moment contribution.
This approach has recently made impressive progress. However, compared to the following options below, the uncertainty is still rather large.
Note, that in the isotope shift (Sec.\,\ref{sec:isoShift}), a large part of this uncertainty cancels. 

The contribution of the individual nucleons $\delta E_{\rm Friar}^N$ is not automatically included by this approach and has to be calculated separately.
}
%
%
The neutron Friar moment is found to be negligible \cite{Pachucki:2015:PRA91_040503}. For the proton, we follow~\cite{NevoDinur:2016:TPE,Friar:2013:PRC88_034004,Miller:2015:lepton-sea} and obtain its value in \MuFourHe{} by using the proton's Friar moment contribution in muonic hydrogen $\delta{}E_{\mathrm{Friar}}^{(p)}(\mu{}\mathrm{H})=0.0247(13)$\,meV provided in~\cite{Carlson:2011:PRA84_020102}.
We scale it with the wavefunction overlap, that depends on the reduced mass ($m_{r}$) and proton number ($Z$) scaling to the third power.
We account for the different number of protons in both systems with an additional $Z$ ratio. 
Another reduced mass scaling factor enters from the third term in Eq.\,(11) of~\cite{Friar:2013:PRC88_034004} according to~\cite{NevoDinur:PC:2016}.
We obtain for the total nucleon Friar moment~\footnote{In Eq.\,(12) of Ref.\,\cite{Krauth:2016:mud}, we used a scaling of the nucleon TPE contribution by the reduced mass ratio to the third power, which is only correct for  $\delta E^N_{\rm inelastic}$. $\delta E^N_{\rm Friar}$ should be scaled with the fourth power \cite{Friar:2013:PRC88_034004,NevoDinur:2016:TPE}. 
This is due to an additional $m_r$ scaling factor compared to the proton polarizability term.
 This mistake has no consequences for $\mu$d yet, as the nuclear uncertainty is much larger, but the correct scaling is relevant for \MuThreeHe and \MuFourHe.
}
\begin{equation}\label{eq:nucleonFriar}
\begin{split}
   \delta{}E_{\mathrm{Friar}}^{N} = & 
  \biggl(\frac{m_r(\mu{}^4\mathrm{He})}
 {m_r(\mu{}\mathrm{H})}
 \frac{Z(\mu{}^4\mathrm{He})}{Z(\mu{}\mathrm{H})}\biggr)^{4} 
   \delta{}E_{\mathrm{Friar}}^{(p)}(\mu{}\mathrm{H})\\[7pt]
                        = &~0.541\pm0.028\mev.
\end{split}
\end{equation}
(the individual terms are provided in footnote \footnote{$m_r(\mu{}\mathrm{H})=185.84\,m_e$, 
$m_r(\mu{}\mathrm{D})=195.74\,m_e$, $m_r(\mu{}^4\mathrm{He})=201.07\,m_e$, $Z(\mu{}\mathrm{H}) = 1$, 
$Z(\mu{}\mathrm{D}) = 1$, $Z(\mu{}^4\mathrm{He}) = 2$, $A(\mu{}\mathrm{D}) = 2$, $A(\mu{}^4\mathrm{He}) = 4$}).\\ 
This value agrees with the result reported in~\cite{Ji:2016:NuclStruc}.
\JI{The total Friar moment contribution according to \textit{option \OBACCA} is then given by the sum of Eqs.\,(\ref{eq:obacca_nuclear}) and (\ref{eq:nucleonFriar})
\begin{equation}
  \delta{}E_{\mathrm{Friar}}^{A+N}(\OBACCA) = 6.68\pm0.31\mev
\end{equation}  
}

\textit{Option \OFRIAR}: The Friar moment contribution can be parameterized as being proportional to the Friar moment $\braket{r^3}_{(2)}$ of the nucleus' electric charge distribution \cite{Friar:1978:Annals}.\\Using $\braket{r^3}_{(2)} = 16.73(10)$\,\mev/fm$^{3}$ \cite{Sick:2014:HeZemach} from measured helium-4 form factors in momentum space, this option yields the most precise value and is furthermore model-independent, as it originates from experimental data. From Eq.\,(43a) in \cite{Friar:1978:Annals} we obtain
\begin{equation}\label{eq:Friar:Friar}
  \delta E_{\rm Friar}^{A+N}(\OFRIAR) = \frac{(Z\alpha)^5 m_r^4}{24} \braket{r^3}_{(2)} = 6.695\pm0.040\,\mev.
\end{equation}
However, expressing the elastic part of the TPE using only the Friar moment does not account for relativistic recoil corrections \cite{Carlson:2017:PC} (see \textit{option \OMART}).

\textit{Option \OBORIE}: The Friar moment contribution can be parameterized proportional to the third power of the nuclear rms charge radius as $C\times \ra^3$, where $C$ is a factor which depends on the model for the radial charge distribution.\\Borie gives a coefficient of $C = 1.40(4)\mev/\fm^3$ (p.\,14 of \cite{Borie:2014:arxiv_v7}). It is valid for a Gaussian charge distribution which is a good assumption for the helium-4 nucleus. The given uncertainty is an estimate of possible deviations from the Gaussian charge distribution \cite{Borie:PC:2017}. For the nuclear charge distribution Borie uses the charge radius $\ra = 1.681(4)\fm$ from Sick \cite{Sick:2008:rad_4He} and obtains
\begin{equation}\label{eq:Friar:rad}
\begin{split}
  \delta E_{\rm Friar}^{A+N}(\OBORIE) =& ~1.40\pm0.04\mev/\fm^3 \times \ra^3\\
 =& ~6.650\pm0.190\mev.
\end{split}
\end{equation}
In addition to this value from Borie, the more recent results of Refs.\,\cite{Ji:2014:FewBodySyst} and \cite{Sick:2014:HeZemach} lead to slightly improved theoretical and experimental values of $C=1.38(3)$ and $C=1.41(2)$, respectively, which are all in agreement with one another \cite{NevoDinur:PC:2018}.
In principle one can benefit from \textit{option \OBORIE} by using the charge radius as a free parameter which will be determined by the measurement of the Lamb shift. This has initially been done in $\mu$p \cite{Pohl:2010:Nature_mup1}. The limiting uncertainty in \MuFourHe, however, comes from the coefficient which is why this option is not attractive until a better value for the coefficient is available.

\textit{Option \OMART}: The Friar moment contribution can be calculated by directly using form factor (FF) parameterizations in momentum space.\\
Martynenko did the calculation for Gaussian and Dipole electric FF parameterizations due to their closed analytical form, following the work of Friar~\cite{Friar:1977:PRC16_1540}. Using a charge radius of $r_\alpha = 1.676(8)$\,fm, Martynenko with Eqs.\,(62) and (63) in \cite{Krutov:2014:JETP120_73} determines an energy contribution of 
\begin{equation}\label{eq:Friar:Martynenko}
  \delta E_{\rm Friar}^{A+N}(\OMART) = 6.61\pm0.07\mev
\end{equation}
for a Gaussian charge distribution. For the less realistic dipole parameterization, Martynenko obtains a value of $7.1958\mev$, which differs by $\sim0.6\mev$ from Eq.\,(\ref{eq:Friar:Martynenko}), illustrating the sensitivity of the Friar moment contribution to the shape of the charge distribution. In his table, however, Martynenko uses the Gaussian charge distribution only.
%
%

\begin{figure}
  \begin{center}
    \includegraphics[angle=0,width=7.cm]{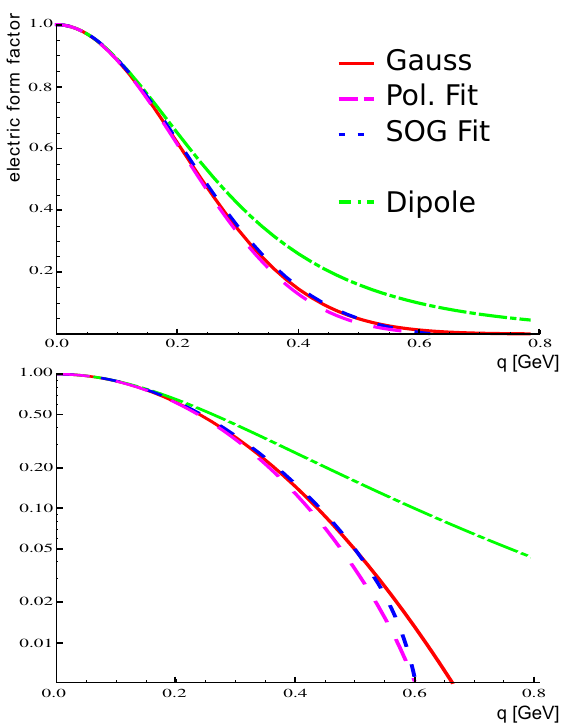}
    \caption{Parameterizations of the alpha particle electric form factor (FF) which are used in the calculations of the Friar moment contribution. Shown are Gaussian (red, solid) and dipole functions (green, dash-dotted) together with two experimentally deduced FF parameterizations (purple and blue, dashed)~\cite{Ottermann:1985:3He_4He_scatt,Sick:2015:PC} on a linear (Top) and logarithmic scale (Bottom). The Gaussian and dipole FF reproduce $r_\alpha = 1.681$\,fm. The experimental fitting results agree with the Gaussian shape, and significantly diverge from the dipole parameterization at momentum transfers $\geq0.2\,\mathrm{GeV}$. We base our analysis on the SOG fit by Sick~\cite{Sick:2015:PC}.}
    \label{fig:FF}
  \end{center}
\end{figure}

\textit{Option \ODATA}: Similar to \textit{option \OMART}, but instead of assuming the FF to be Gaussian we use FFs based on measured data.\\
Sick provided us with an improved parameterization of the charge distribution from current world data on elastic electron scattering on \FourHe{} by means of a sum of Gaussians charge distribution~\cite{Sick:2015:PC}. By numerical Fourier transformation we obtain the electric FF which is used in Eqs.\,(62) and (63) in \cite{Krutov:2014:JETP120_73}. In Fig.\,\ref{fig:FF} we compare the FF obtained with the parameterization from Sick, with other parameterizations.
With the FF parameterization from Sick we obtain a Friar moment contribution of
\begin{equation}\label{eq:Friar:FF}
     \delta{}E_{\rm Friar}^{A+N}(\ODATA) = 6.65\pm0.09\mev.
\end{equation}
This value is in agreement with the value reported by Martynenko (\textit{option \OMART}) for a simple Gaussian FF. The value in Eq.\,(\ref{eq:Friar:FF}) has a slightly larger uncertainty than the one reported in Eq.\,(\ref{eq:Friar:Martynenko}). The advantage of \textit{option \ODATA}, however, is its model-independence due to the experimentally measured FF. In contrast to \textit{option \OFRIAR}, the value in Eq.\,(\ref{eq:Friar:FF}) accounts for relativistic recoil corrections. 
\\\\
All five options presented here are in agreement, whereas their uncertainties differ a lot. 
Since the dominating uncertainty arises from the inelastic contributions and \textit{not} from the Friar moment contribution, the different options do not influence the total TPE contribution significantly. 
The best value to use might be \textit{option \ODATA} for the reasons discussed above. This choice is also recommended by Carlson \cite{Carlson:2017:PC}. Note, that in the case of muonic helium-3 ions, less data is available, which is the reason why for \MuThreeHe \cite{Franke:2016:mu3HeTheo}, \textit{option \ODATA} was not considered.
  However, in order to have a consistent treatment for elastic and inelastic contributions one might also consider \textit{option \OBACCA}. The consistency is then given because the inelastic calculation discussed in the next section does not include relativistic corrections either. There are reasons to expect that the (not calculated) relativistic corrections of both terms cancel each other.
  Since the choice does not significantly influence the total uncertainty of the extracted charge radius, we decide to use \textit{option \ODATA}, acknowledging that the other choice is also valid.



From \textit{option \ODATA} we obtain as Friar moment contribution
\begin{equation}\label{eq:Friar:FriarTotal}
  \delta E_{\rm Friar}^{A+N} = \delta{}E_{\rm Friar}^{A+N}(\ODATA) = 6.65 \pm 0.09 \mev.
\end{equation} 
\\\\

\subsection{\MuFourHe{} Polarizability}
\label{sec:pol}

The second component of the two-photon exchange in Eq.\,(\ref{eq:TPE}) is given by the inelastic nuclear ``polarizability'' contribution, $\delta{}E_{\rm inelastic}^{A}$, that stems from the virtual excitation of the nucleus in the two-photon interaction (see Fig.~\ref{fig:feynman2}\,(c)+(d)).
%
%
The initial calculation of the polarizability was done by Joachain \cite{Joachain:1961:pol} in 1961. Rinker \cite{Rinker:1976:he_pol} in 1976 and Friar \cite{Friar:1977:PRC16_1540} in 1977 improved the calculation and obtained a value of 3.1(6)\,meV~\cite{Friar:1977:PRC16_1540}.
Martynenko used this value in his summary\,\cite{Krutov:2014:JETP120_73}, as did Borie in previous versions of hers\,\cite{Borie:2014:arxiv_v7}.
Recently, a more accurate calculation of the \MuFourHe{} nuclear polarizability was done by Ji~\etal~\cite{Ji:2013:PRL111,Ji:2018:nuclPOL} using two parameterizations of the nuclear potential.
Their calculation uses the AV18 nucleon-nucleon (NN) force plus the UIX three-nucleon (NNN) force, as well as NN forces plus NNN forces from chiral effective field theory ($\chi$EFT) to calculate the terms up to the order $(Z\alpha{})^5$.
It provides an energy contribution of 2.47(15)\,meV that is in agreement with Friar's value
but four times more precise.
Borie adopted the value of Ji \etal~\cite{Ji:2013:PRL111} in the newest version of 
her summary. 
%
%

%
In muonic deuterium, Pachucki \cite{Pachucki:2011:PRL106_193007} found that the elastic part is exactly canceled by a part of the inelastic. These terms are called $\delta{}_{Z1}^{1}$ and $\delta{}_{Z3}^{1}$ in Ji \etal~\cite{Ji:2013:PRL111}. 
We assumed this cancellation to be exact in our muonic deuterium theory 
summary~\cite{Krauth:2016:mud}.
Here we treat both parts of the TPE separately and do not use the cancellation.
%
%
For the nuclear polarizability contribution we adopt the most recent value
\JI{\cite{Ji:2018:nuclPOL}
\begin{equation}\label{eq:nuclearPOL}
  \delta{}E_{\mathrm{inelastic}}^{A}= 2.35\pm0.13\,\mathrm{meV}.
\end{equation}
}
Next, we account for the contribution due to the polarizability of the individual nucleons $\delta E_{\rm inelastic}^N$. 
\JI{In \cite{Hernandez:2016:POLupdate}, the TRIUMF/Hebrew group provides a value which was later updated to \cite{Ji:2018:nuclPOL} 
\begin{equation}\label{eq:Npol:Bacca}
  \delta E_{\rm inelastic}^N({\rm Hernandez}) = 0.34\pm0.20\mev.
\end{equation}
}
This value is obtained by scaling the contribution for a single proton of 0.0093(11)\mev{}~\footnote{\label{note1}The contribution for a single proton is the sum of an inelastic term $0.0135\mev$ \cite{Carlson:2014:PRA89_022504} and a subtraction term $\delta^{p}_{\rm subtr} = -0.0042(10)\mev$ \cite{BirseMcGovern:2012}.} 
by the number of protons and neutrons~\footnote{\label{note2}Assuming isospin symmetry, the value of the neutron polarizability contribution is the same as the one of the proton, but, as in \cite{NevoDinur:2016:TPE}, an additional uncertainty of 20\% is added, motivated by studies of the nucleon polarizabilities \cite{Myers:2014:comptonScatt}.}, as well as with the wavefunction overlap, according to Eq.\,(19) of Ref.\,\cite{NevoDinur:2016:TPE}. It also includes a 29\% correction for estimated medium effects and possible nucleon-nucleon interferences.
A smaller uncertainty for $\delta E_{\rm inelastic}^N$ could be achieved starting with the inelastic contribution for a proton-neutron pair in muonic deuterium, which was determined by 
Carlson \etal\ to be $\delta{}E_{\mu{}\mathrm{D}}^{\mathrm{hadr}}=0.028(2)$\,meV~\cite{Carlson:2014:PRA89_022504}. We scale Carlson's value with
%
the nucleon number $A$ and the wavefunction overlap with the nucleus to obtain
\begin{equation}\label{eq:nucleonPOL}
\begin{split}
  \delta{}E_{\mathrm{\rm \mu4He^+}}^{\rm hadr} = &~\frac{A(\mu{}^4\mathrm{He})}{A(\mu{}\mathrm{D})}\biggl(\frac{m_r(\mu{}^4\mathrm{He})} {m_r(\mu{}\mathrm{D})}\frac{Z(\mu{}^4\mathrm{He})}{Z(\mu{}\mathrm{D})}\biggr)^{3} \delta{}E_{\mu{}\mathrm{D}}^{\mathrm{hadr}} \\[7pt]
                         = &~0.486\pm0.069\,\text{meV}.
\end{split}
\end{equation}
The uncertainty of this value was chosen a factor of two larger than the one
that would be obtained by scaling the \MuD{} value.
This accounts
for possible shadow effects that could exist in the \FourHe{} nucleus.
This uncertainty estimate has been confirmed by Bacca, Carlson and Gorchtein~\cite{Gorchtein:PC:2015},
until better results for \MuFourHe{} and \MuThreeHe{} are available.

We have to add to Eq.\,(\ref{eq:nucleonPOL}) the contribution due to the subtraction term for protons and neutrons (see footnotes~\textsuperscript{\ref{note1}} and \textsuperscript{\ref{note2}}) and obtain by scaling from $\mu$H
%
%
%
%
%
%
\begin{equation}\label{eq:subtr}
\begin{split}
   \delta{}E_{\mathrm{sub}} = &
   ~\biggl(\frac{m_r(\mu{}^4\mathrm{He})}
  {m_r(\mu{}\mathrm{H})}\frac{Z(\mu{}^4\mathrm{He})}{Z(\mu{}\mathrm{H})}\biggr)^{3}\\[7pt]
  & \times{}2(\delta^{p}_{\rm subtr}+\delta^{n}_{\rm subtr})\\[7pt]
                         = &~-0.170\pm0.032\,\text{meV}.
\end{split}
\end{equation}
The sum of Eqs.\,(\ref{eq:nucleonPOL}) and (\ref{eq:subtr}) yields an alternative value to Eq.\,(\ref{eq:Npol:Bacca}) of 
\begin{equation}\label{eq:Npol:carlson}
  \delta E_{\rm inelastic}^N({\rm Carlson}) = 0.316\pm0.076\mev
\end{equation}
which is in good agreement with Eq.\,(\ref{eq:Npol:Bacca}) but three times more precise.

Summarizing, the nucleon polarizability can be scaled either from the proton, or from the deuteron. It is not clear which option is the better one. We therefore decided to average Eq.\,(\ref{eq:Npol:Bacca}) and (\ref{eq:Npol:carlson}), which yields
\JI{
\begin{equation}\label{eq:Npol}
  \delta E^N_{\rm inelastic} = 0.33 \pm 0.20 \mev.
\end{equation} 
}
\\\\

%
%
%
%
%

\JI{
For the total TPE contribution, which is the sum of the Friar moment contribution (Eq.\,(\ref{eq:Friar:FriarTotal})) and the nuclear and the nucleon polarizability contributions (Eqs.\,(\ref{eq:nuclearPOL}),\,(\ref{eq:Npol})) we obtain
\begin{equation}\label{eq:tpe:total}
  \Delta E^{\rm LS}_{\rm TPE} = 9.34 \pm 0.25\mev.
\end{equation}

Here, the nuclear and nucleon polarizability contribute 0.11\mev{} and 0.20\mev{} to the uncertainty, respectively.

The value of Eq.\,(\ref{eq:tpe:total}) agrees well with the TPE contribution of 9.37(45)\mev{} obtained through \textit{ab initio} calculations \cite{Ji:2018:nuclPOL}.
}
A dispersive approach as given for helium-3 \cite{Carlson:2016:tpe} is required also for helium-4 in order to crosscheck the value provided here.
The uncertainty of the total TPE contribution is about 4 times larger than the value given
as uncertainty goal due to the achieved experimental precision. An improvement of this value will directly improve the extraction of the charge radius.

\section{\MuFourHe{} Fine Structure}
\label{sec:fs}
The \TwoP{} fine structure splitting (FS) has been calculated by Borie~\cite{Borie:2014:arxiv_v7},
Martynenko (Elekina~\etal\ \cite{Elekina:2011:mu4He}) (see Tab.~\ref{tab:theory0}). 
Both determinations agree within 0.020\mev.
The leading order contribution to Borie's fine structure is given by the Dirac
term of the order $(Z\alpha{})^4$ (\#f1).
Borie provides additional recoil corrections to her Dirac value
(\#f2) not covered by her relativistic Dirac wavefunction approach.
These corrections are already included in Martynenko's term (\#f3). 
Martynenko separately calculates corrections to his term like the $(Z\alpha{})^6$ contribution 
(\#f4a), as well as an additional correction of the order $(Z\alpha{})^6m_1/m_2$ (\#f4b).
The latter term is new in our FS summary and was not accounted for
in $\mu{}$D~\cite{Krauth:2016:mud} due to its negligible size.
We sum the Dirac contributions of both authors, including all given corrections.
Comparing this sum, we find an unexpected 
difference of 0.025\mev{} in the Dirac-term calculation between both authors, as opposed to the much better agreement in the leading order Uehling term of the Lamb shift. As our choice, we decided to use the value of Martynenko, which is in agreement with a very recent calculation of Korzinin \etal{} \cite{Korzinin:2018:lamb} (see \textit{Note added in proof} at the end of this work).
%
%

%
Further corrections to the FS are given by eVP insertions in the FS interaction.
Borie, Martynenko and Karshenboim have calculated the 1-loop eVP in
both, one- (\#f5a) and two Coulomb lines (\#f5b) and 
get matching results for the sum of both terms.
Only Martynenko calculated the two-loop K\"all\'en-Sabry-type diagrams (\#f6a) (corresponding to Fig.\,\ref{fig:feynman}, \#4 in the Lamb shift).
The consecutive two-loop correction in two Coulomb lines (\#f6b) 
has been calculated by Martynenko, Borie, and Karshenboim. We use the value from Karshenboim, as they included some higher order terms as well. 
The corrections of the order $\alpha{}^2(Z\alpha{})^4m$ (\#f7) are only calculated 
by Karshenboim and are included in our summary.
The correction of order $\alpha(Z\alpha)^6$ (\#f11*) is only provided by Martynenko whose value we adopt.
Martynenko also calculates the value of the one loop $\mu$VP contribution to
the FS that we adopt for our summary (\#f12*).
Contributions of the muon anomalous magnetic moment to the fine 
structure were provided by Borie and Martynenko and agree perfectly (\#f8,\#f9).
The first order finite size contribution has a minor influence on the 
\TwoP{} FS interval because the $2P_{1/2}$ level has a small, yet non-zero wavefunction at the origin.
Calculations of the first order term (\#f10a) by Borie and Martynenko agree very 
well and we use the average.
The term also appears as nuclear structure dependent part of the Lamb shift (\#r8)
and has to be accounted for in both the Lamb shift and the FS.
The radius dependence is neglected in the FS since it only provides a minor influence
to the total energy difference.
For the second order contribution (\#f10b) we adopt the only available
calculation by Martynenko.
Using the named values, the total \TwoP{} FS in \MuFourHe{} is given by
\begin{equation}\label{eq:fs:tot}
  \Delta{}E_{2P_{3/2}-2P_{1/2}}= 146.1828\pm0.0003\,\text{meV}.
\end{equation}
This is in good agreement with the published value of Martynenko \cite{Elekina:2011:mu4He}, and disagrees with the value from Borie \cite{Borie:2014:arxiv_v7}. The reason is the difference of the leading order Dirac term between those two.
The given uncertainty of the total fine structure arises due to negligible inconsistencies between Martynenko and Borie, which are not clarified yet.

\section{Summary for \MuFourHe{}}
\label{sec:summary}

We provided a summary of the Lamb shift and \TwoP{}-fine structure in \MuFourHe{} that
will be used for the extraction of the alpha particle charge radius from
the measurements performed at PSI~\cite{Nebel:2012:LEAP_muHe,Antognini:2011:Conf:PSAS2010}.
We compared the calculations of Borie~\cite{Borie:2014:arxiv_v7},
the Martynenko group~\cite{Krutov:2014:JETP120_73,Elekina:2011:mu4He}, the Karshenboim group~\cite{Karshenboim:2012:PRA85_032509,Korzinin:2013:PRD88_125019},
the Jentschura group~\cite{Jentschura:2011:SemiAnalytic}, and the TRIUMF/Hebrew group\,\cite{Ji:2013:PRL111,Ji:2014:FewBodySyst,Hernandez:2016:POLupdate,Ji:2018:nuclPOL}.
After sorting and comparing all individual terms we found some
discrepancies between the different sources (see Tabs~\ref{tab:theory1}-\ref{tab:theory0}):
Two-loop eVP (\#4,\#5), $\alpha^2(Z\alpha)^4m$ contribution (\#29) and
higher orders (\#12-\#21) in the radius-independent Lamb shift contributions, different finite size contributions (\#r1,\#r3,\#r3',\#r4,\#r5),
the elastic Friar moment contribution (Eq.\,(\ref{eq:Friar:Friar}) and following), as well as the sum of the Dirac contributions in the FS (\#f1-f4).
Several contributions are only single-authored  (\#r2',\#r2b',\#r6,\#r7,\#r8, and others). In order to have a reliable theory prediction of the Lamb shift we encourage the theory groups to perform independent calculations to crosscheck the terms calculated by others. 
\JI{
In summary, we obtain the total energy difference of the $\TwoS_{1/2}\rightarrow\TwoP_{1/2}$ Lamb shift transition in \MuFourHe{} as a function of the nuclear charge radius \ra{} as
\begin{widetext}
\begin{align}
  \Delta{}E_{\rm (2P_{1/2}-2S_{1/2})} &= \Delta{}E_{(QED+Recoil)}~+~\Delta{}E_{(Finite~Size)}~+~\Delta{}E_{\rm TPE}^{\rm LS}\\[4pt]
&= 1668.489(14)~\mev\nonumber\\
     &~\hspace{82pt}~-~106.354(8)~\mev/\fm^2\times{}\ra^2 +  0.078(11)\mev\label{eq:theo:totalterms}\\[4pt]
              &~\hspace{165pt}~+~9.340(250)~\mev\nonumber\\
&= 1677.907(251) - 106.354(8)\mev/\fm^2\times\ra^2.\label{eq:theory:totalsum}\\[5pt]
(\text{for comparison}&\text{ use:}~\ra^2\approx{}2.83\,\fm^2)\nonumber
\end{align}
\end{widetext}
%
%
%

%
\vspace{5mm}
The $\Delta{}E_{\rm (2P_{1/2}-2S_{1/2})}$ energy difference corresponds to the sum of Eqs.\,(\ref{eq:LS:tot}), (\ref{eq:finsize:tot}), and (\ref{eq:tpe:total}).
The $\Delta{}E_{\rm (2P_{3/2}-2S_{1/2})}$ energy difference is obtained by including the fine structure from Eq.\,(\ref{eq:fs:tot}) which yields
\begin{equation}
\begin{split}
  \Delta{}E_{\rm (2P_{3/2}-2S_{1/2})} &= 1824.090(251)\\ &\hphantom{=~}- 106.354(8)\mev/\fm^2\times\,\ra^2.
\end{split}
\end{equation}
}
The currently limiting factor in the \MuFourHe{} theory originates from the two-photon exchange (TPE) contribution, where mainly the uncertainty of the inelastic nucleon polarizability contribution (Eq.\,(\ref{eq:Npol})) is dominating. Improving the terms which constitute the TPE will directly improve the value of the alpha particle charge radius determination.  
Using the prediction of the Lamb shift from the compiled theory of this work, we derive a theory uncertainty of the \FourHe{} nuclear charge radius from the laser spectroscopy measurement in muonic helium-4 ions of $<0.0008\,\mathrm{fm}$. Compared to this value the expected experimental uncertainty will be small. Hence, with the theory presented in this compilation and the to-be-published measurement, we expect an improvement of the previous best value by a factor of $\sim5$.


\section{The \textsuperscript{3}He --\textsuperscript{4}He isotope shift}\label{sec:isoShift}

The "isotope shift" (IS) refers to the 
\RP{
change of a transition frequency between different isotopes
of the same element.
It originates mainly from the nuclear {\em mass} difference and
the change of {\em charge radii}.
Since the masses of the lightest nuclei are very well
known~\cite{Heisse:2017:Mass_p,Zafonte:2015:Mass_D_T,Myers:2013:Mass_T_3He,VanDyck:2004:Mass_4He,Brodeur:2012:68He,Blaum:2006:Masses,Mohr:2016:CODATA14}, and theory of the
IS is simplified by beneficial
cancellations~\cite{Jentschura:2011:IsoShift,CancioPastor:2012:PRL108},
a measurement of the same transition in two isotopes can be used to
obtain an accurate value of the
squared charge radius difference, 
e.g.\ $\rh^2 - \ra^2$ for $^3$He and $^4$He
\cite{Shiner:1995:heliumSpec,Rooij:2011:HeSpectroscopy,CancioPastor:2012:PRL108,Zheng:2017:He_Iso}.
}

For the CREMA measurements in muonic $^3$He and $^4$He, one could calculate the IS from the {\em absolute} charge radii, determined using Eq.\,(18) in \cite{Franke:2016:mu3HeTheo} and Eq.\,(\ref{eq:theo:totalterms}) in here. The accuracy of both muonic radii is limited by the uncertainty in the nuclear and nucleon two-photon exchange (TPE) contribution. However, the uncertainties due to nuclear model-dependence and the ones due to the single nucleons, respectively, are strongly correlated between the two isotopes.

Here we attempt to reduce the uncertainty in the theory of the TPE contributions to the IS in muonic $^3$He and $^4$He, taking into account these correlations. 

For both helium isotopes ($i=h,\alpha$), the transition energy is given in the form
\begin{equation}
  \Delta E_{\rm LS}^{(i)} = E_{r-\rm ind.}^{(i)}+ c_i r_i^2 + E_{\rm TPE}^{(i)}\label{eq:LS:theo},
\end{equation}
where $E_{r-\rm ind.}^{(i)}$ is the radius-independent QED part of the Lamb shift, the second term is the radius-dependent part of the Lamb shift with the charge radius $r_i$ and the coefficient $c_i$, and $E_{\rm TPE}^{(i)}$ is the two-photon exchange. Using the measured transition energy $h\nu^{(i)}_{\rm LS}$ of isotope $i$ and Eq.\,(\ref{eq:LS:theo}), the square of the charge radius $r_i$ is
\begin{equation}
  r_i^2 = \frac{1}{c_i} \left(h\nu^{(i)}_{\rm LS} - E_{r-\rm ind.}^{(i)} - E_{\rm TPE}^{(i)}\right).
\end{equation}

The value of each charge radius will be limited by the theory uncertainty from the two-photon exchange (TPE) contribution. In order to exploit correlations, we use for the IS the TPE results from the TRIUMF/Hebrew group \cite{Ji:2013:PRL111,NevoDinur:2016:TPE,Hernandez:2016:POLupdate,Ji:2018:nuclPOL} alone. The TRIUMF/Hebrew group has consistently calculated the TPE for both, \MuFourHe{} and \MuThreeHe{}. This means using \textit{option \OBACCA} for the Friar moment contribution, described in Sec.\,\ref{sec:friar}, which is not what we chose for the extraction of the alpha charge radius. However, due to correlations in the IS, the large uncertainty of \textit{option \OBACCA} partly cancels.

The uncertainty of the charge radius $r_i$ is obtained by propagating the uncertainties from the experimental value $h\nu_{\rm LS}^{(i)}$~\footnote{To be published. The experimental uncertainty will be dominated by statistics.} and the theory values $E_{r-\rm ind.}^{(i)}$, $c_i$, and $E_{\rm TPE}^{(i)}$~\footnote{The uncertainty in the TPE contribution dominates by far the total uncertainty in the charge radius.}. 
The isotope shift is then given by
\begin{align}
  r_h^2 - r_\alpha^2 &= \frac{h\nu^{(h)}_{\rm LS} - E_{r-\rm ind.}^{(h)} - E_{\rm TPE}^{(h)}}{c_h} \nonumber\\
                   &  \hphantom{=~} - \frac{ h\nu^{(\alpha)}_{\rm LS} - E_{r-\rm ind.}^{(\alpha)} - E_{\rm TPE}^{(\alpha)}}{c_\alpha}\\
  &=   \frac{h\nu^{(h)}_{\rm LS}}{c_h} - \frac{h\nu^{(\alpha)}_{\rm LS}}{c_\alpha}
      - \left(\frac{E_{r-\rm ind.}^{(h)}}{c_h}- \frac{E_{r-\rm ind.}^{(\alpha)}}{c_\alpha}\right)\nonumber\\
  &\hphantom{=~}    -\left(\frac{E_{\rm TPE}^{(h)}}{c_h}- \frac{E_{\rm TPE}^{(\alpha)}}{c_\alpha}\right)\label{eq:IsoShift}\\
  &= \Delta_\nu - \Delta_{r-\rm ind.} - \Delta_{\rm TPE},\label{eq:IsoShift:short}
\end{align}
where $\Delta_\nu$ contains the experimental Lamb shift transition energies ($h\nu_{\rm LS}^{(h)}$ and $h\nu_{\rm LS}^{(\alpha)}$). All other contributions including their uncertainties are listed in Tab.\,\ref{tab:iso_terms}. A simple Gaussian propagation of these uncertainties is only correct for uncorrelated terms. In the following we discuss the correlations in the uncertainties of the TPE terms and 
show how to get rid of the uncertainty correlations which arise due to nuclear modeling and due to scaling the nucleon contributions.
 
\begin{table}
\begin{center}
{\renewcommand{\arraystretch}{1.2}
\begin{tabular}{|l f{2} f{4} c|}
\hline
                        & \multicolumn{2}{c}{value} & \multicolumn{1}{c|}{source}\\
\hline
\hline
$E_{r-\rm ind.}^{(h)}$      & 1644.482     & \pm0.015       & \cite{Franke:2016:mu3HeTheo}, Eq.\,(18), first two terms\\
$E_{r-\rm ind.}^{(\alpha)}$  & 1668.567     & \pm0.018       & Eq.\,(\ref{eq:theo:totalterms}), 1st and 3rd term\\
$c_h$                   & -103.518      & \pm0.010       & \cite{Franke:2016:mu3HeTheo}, Eq.\,(18), 3rd term\\
$c_\alpha$               & -106.354      & \pm0.008       & Eq.\,(\ref{eq:theo:totalterms}), 2nd term\\
$E_{\rm TPE}^{(h)}$        &  15.49       & \pm0.32        & \cite{Ji:2018:nuclPOL}, Tab.\,7\\
$E_{\rm TPE}^{(\alpha)}$    &   9.37       & \pm0.45        & \cite{Ji:2018:nuclPOL}, Tab.\,7\\
\hline
\end{tabular}
}
\caption{The values and uncertainties of the terms which appear in the isotope shift (Eq.\,(\ref{eq:IsoShift})). 
The dominating uncertainties arise from the TPE terms of both helium isotopes. Note that their uncertainties are correlated and should not simply be propagated for the isotope shift. For a detailed discussion, see text. The values are given in units of \mev.}
\label{tab:iso_terms}
\end{center}
\end{table}

Since the uncertainties in the TPE terms are by far the dominating uncertainty in the extraction of the isotope shift, we restrict the discussion of correlations to the TPE contributions only and have a closer look into the TPE term $\Delta_{\rm TPE}$ from Eq.\,(\ref{eq:IsoShift:short})
\begin{equation}
\begin{split}\label{eq:IsoTPE}
\Delta_{\rm TPE} &= \frac{E_{\rm TPE}^{(h)}}{c_h} - \frac{E_{\rm TPE}^{(\alpha)}}{c_\alpha}\\
               &= \frac{\delta E^{A, (h)}_{\rm Friar} + \delta E^{A, (h)}_{\rm inel.} + a_h\delta E^{(p)}_{\rm Friar} + b_h\delta E^{(p)}_{\rm inel.}}{c_h}\\
               & \hphantom{=~}- \frac{\delta E^{A, (\alpha)}_{\rm Friar} + \delta E^{A, (\alpha)}_{\rm inel.} + a_\alpha\delta E^{(p)}_{\rm Friar} + b_\alpha\delta E^{(p)}_{\rm inel.}}{c_\alpha}\\
               &= \frac{\delta E^{A, (h)}_{\rm TPE}}{c_h} - \frac{\delta E^{A, (\alpha)}_{\rm TPE}}{c_\alpha}\\
               & \hphantom{=~}+ (\frac{a_h}{c_h}-\frac{a_\alpha}{c_\alpha})\delta E^{(p)}_{\rm Friar} + (\frac{b_h}{c_h}-\frac{b_\alpha}{c_\alpha})\delta E^{(p)}_{\rm inel.},\\
\end{split}
\end{equation}
where, following Eq.\,(\ref{eq:TPE}), we break down the TPE contribution into its constituents and explicitly write the nucleon terms as a scaling factor $a_{h/\alpha}$, $b_{h/\alpha}$ times the respective contribution from the proton. The values of the TPE terms and the scaling factors in Eq.\,(\ref{eq:IsoTPE}) are listed in Tab.\,\ref{tab:tpe_terms}. For the scaling factors $a_{h/\alpha}$ and $b_{h/\alpha}$ compare Eqs.\,(17) and (19) in \cite{NevoDinur:2016:TPE}. In the last line we sum up the nuclear terms for both isotopes, respectively, and we reorder the nucleon terms to make the cancellations visible which appear between the scaling factors.

\begin{table*}
\begin{center}
{\renewcommand{\arraystretch}{1.2}
\begin{tabular}{|l f{9} f{6} f{6} f{7} f{5} f{10}|}
\hline
                                & \multicolumn{3}{c}{$i=h$} & \multicolumn{3}{c|}{$i=\alpha$}\\
\hline
\hline
                                & \multicolumn{1}{c}{AV18+UIX\,\cite{NevoDinur:2016:TPE}}
                                             &\multicolumn{1}{c}{$\chi$EFT\,\cite{NevoDinur:2016:TPE}}
                                                         &\multicolumn{1}{c}{avg.\,\cite{Hernandez:2016:POLupdate}}
                                                                     &\multicolumn{1}{c}{AV18+UIX\,\cite{Ji:2013:PRL111}}
                                                                             &\multicolumn{1}{c}{$\chi$EFT\,\cite{Ji:2013:PRL111}}
                                                                                     &\multicolumn{1}{c|}{avg.\,\cite{Hernandez:2016:POLupdate}~~}\\
\hline
$\delta E^{A,(i)}_{\rm Friar}$ [meV]& 10.356 & 10.618 & 10.49(24)[{19\atop16}] & 5.936 & 6.337 & 6.14(31)[{28\atop12}]^*\\
$\delta E^{A,(i)}_{\rm inel.}$ [meV]&  4.21^* &  4.25^* &  4.23(12)[{03\atop12}]^* & 2.29^* & 2.42^* & 2.36(11)[{09\atop06}]^*\\
\hline
Sum ($\delta E^{A,(i)}_{\rm TPE}$) & 14.56      & 14.87    & 14.72(29)[{21\atop20}] & 8.23  & 8.76 & 8.49(40)[{37\atop13}]\\
\hline
\hline
$a_i$                       & \multicolumn{3}{c}{21.1511} & \multicolumn{3}{c|}{21.9247}\\
$b_i$                       & \multicolumn{3}{c}{29.5884} & \multicolumn{3}{c|}{40.5284}\\
$\delta E^{(p)}_{\rm Friar}$ [meV] & \multicolumn{6}{c|}{0.0247(13)}\\
$\delta E^{(p)}_{\rm inel.}$ [meV] & \multicolumn{6}{c|}{0.0093(11)}\\
\hline
\end{tabular}
}
\caption{Values and uncertainties of the TPE terms which appear in the term $\Delta_{\rm TPE}$ (Eq.\,(\ref{eq:IsoTPE})). Entries, labeled with $^*$ are updated through \cite{Ji:PC:2018,Ji:2018:nuclPOL}. The upper part represents the nuclear and the lower part the nucleon terms. The nuclear terms are calculated using the AV18+UIX nuclear potential and $\chi$EFT. The values from the two approaches are used in order to determine the uncertainty due to nuclear model-dependence. The values given under ``avg.'' are the ones which are then used to infer a value of the IS calculation. The total uncertainties are given in the first brackets. The individual uncertainty contributions are shown in the squared brackets. Uncertainties due to nuclear model dependence are given by the top value, all other uncertainties are summarized in the bottom value.
\RP{The $a_i$ and $b_i$ denote the scaling factors for obtaining the nucleon 
Friar moment contributions from the value calculated for muonic hydrogen, 
see text.
}
}
\label{tab:tpe_terms}
\end{center}
\end{table*}

We discuss first the nuclear uncertainties and then the nucleon part.

\textit{The nuclear part.} The nuclear TPE contribution (nuclear Friar moment + nuclear polarizability) for both muonic helium isotopes has been calculated by the TRIUMF/Hebrew group \cite{Ji:2013:PRL111,NevoDinur:2016:TPE,Hernandez:2016:POLupdate,Ji:2018:nuclPOL} using the AV18 potential and using $\chi$EFT \footnote{For the Friar moment, the calculation of the TRIUMF/Hebrew group corresponds to \textit{option \OBACCA}, discussed in Sec.\,\ref{sec:tpe}.} The difference between the numbers of the two calculations serves as estimate for the uncertainty due to nuclear model-dependence. Using the same method for the isotope shift in Eq.\,(\ref{eq:IsoTPE}), i.e.\ inserting the different values given in Tab.\,\ref{tab:tpe_terms} one after the other, a good estimate for the nuclear model uncertainty is obtained. \JI{It amounts to 0.0010\mev}. 

\JI{The TRIUMF/Hebrew group also provides an uncertainty due to sources other than the nuclear model which are detailed in the supplementary material of \cite{NevoDinur:2016:TPE}. In Eq.\,(16) therein, this uncertainty was only given for muonic helium-3 ions. An updated value for helium-3 and a first value for helium-4 ions was later provided in \cite{Ji:PC:2018}. These values amount to 0.20\mev and 0.13\mev, respectively, see also Tab.\,\ref{tab:tpe_terms}. A propagation of these two uncertainties via Eq.\,(\ref{eq:IsoTPE}) leads to 0.0023\mev. 
}

\textit{The nucleon part.} Similar to the nuclear part, the nucleon TPE contribution for the two isotopes consists of the nucleon Friar moment contribution and the nucleon polarizability contribution. 

The nucleon Friar moment is obtained using the Friar moment from muonic hydrogen of $\delta E^{(p)}_{\rm Friar} = 0.0247(13)\mev$ \cite{Carlson:2011:PRA84_020102}. This value is multiplied with a scaling factor as it is done in Eq.\,(\ref{eq:IsoTPE}). The scaling factor for \MuThreeHe{} is $a_h=21.1511$, for \MuFourHe{} $a_\alpha=21.9247$. Propagating the uncertainty of 0.0013\mev{} via Eq.\,(\ref{eq:IsoTPE}) leads to $2\times10^{-6}\mev$, which is negligible.

The nucleon polarizability contribution is scaled from the single proton polarizability which (including the subtraction term) amounts to $\delta E^{(p)}_{\rm inel.} = 0.0093(11)\mev{}$ (see text below Eq.\,(\ref{eq:Npol:Bacca})). The scaling works according to Eq.\,(23) in \cite{Miller:2015:lepton-sea} which results in scaling factors of $b_h=29.5884$ and $b_\alpha=40.5284$ for muonic helium-3 and -4, respectively. Propagating the uncertainty from the nucleon polarizability of 0.0011\mev{} via Eq.\,(\ref{eq:IsoTPE}) we obtain an uncertainty for the isotope shift of 0.0001\mev{}.

\JI{\textit{The total TPE uncertainty.} In total, the TPE contributions to the isotope shift lead to an uncertainty of 0.0025\mev, 
} which results from the uncertainties added in quadrature. It is dominated by the above discussed uncertainty from the nuclear parts.\\\\

The uncertainties due to the radius-independent QED terms $E_{r-\rm ind.}^{(i)}$ and the coefficients $c_i$ are used as presented in Tab.\,\ref{tab:iso_terms} and propagated via Gaussian propagation of uncertainties through Eq.\,(\ref{eq:IsoShift}). We obtain isotope shift uncertainties of 0.0002\mev{} and 0.0004\mev{} from the radius-independent terms and the coefficients, respectively.\\\\

\JI{Inserting the numbers from Tab.\,\ref{tab:iso_terms} into Eq.\,(\ref{eq:IsoShift}), the $\rh^2-\ra^2$ isotope shift from muonic helium spectroscopy is then given by
\begin{equation}\label{eq:iso:final}
\begin{split}
  \rh^2-\ra^2 &= \left(\frac{h\nu_{\rm LS}^{(h)}/\mev}{-103.518} - \frac{h\nu_{\rm LS}^{(\alpha)}/\mev}{-106.354}\right)\fm^2\\
             &\hphantom{=~} + 0.1970\fm^2 + 0.0615\fm^2\\
             &\hphantom{=~} (\pm 0.0002^{\rm QED} \pm 0.0004^{\rm coeff.} \pm 0.0025^{\rm TPE})\fm^2\\[4pt]
  &= \left(\frac{h\nu_{\rm LS}^{(\alpha)}/\mev}{106.354} - \frac{h\nu_{\rm LS}^{(h)}/\mev}{103.518}\right)\fm^2\\
             &\hphantom{=~} + 0.2585\fm^2 \pm 0.0025^{\rm theo} \fm^2.
\end{split}
\end{equation}
}
Note that without taking into account the correlations and using instead the TPE contributions from Eq.\,(17) in Ref.\,\cite{Franke:2016:mu3HeTheo} and from Eq.\,(\ref{eq:tpe:total}) in this work, the last line of Eq.\,(\ref{eq:iso:final}) would read $0.2570(56)\fm^2$, which is in good agreement, but with a twice larger uncertainty.

Since the experimental uncertainty is expected to be small compared to the theory uncertainty of $0.0025\fm^2$, the extraction of the isotope shift from muonic helium ions will compete with the uncertainty from previous measurements \cite{Shiner:1995:heliumSpec,CancioPastor:2012:PRL108,Rooij:2011:HeSpectroscopy,Zheng:2017:He_Iso} and may shed new light on the discrepancy between those.

\section{Conclusion}\label{sec:conclusion}

In the first part of this work we have summarized and compiled all available theory contributions to the \TwoSTwoP{} Lamb shift in \MuFourHe, which is necessary in order to extract the alpha charge radius from laser spectroscopy measurements in muonic helium-4 ions. The result of our compilation is shown in Eq.\,(\ref{eq:theo:totalterms}).

In the second part, we studied the theory uncertainties which enter the value of the \textsuperscript{3}He--\textsuperscript{4}He isotope shift that can be extracted from the CREMA measurement in \MuThreeHe and \MuFourHe. \JI{We obtain a total theory uncertainty of $0.0025\fm^2$, see Eq.\,(\ref{eq:iso:final})}. This uncertainty is much smaller than a discrepancy between previous isotope shift measurements in electronic helium atoms \cite{Shiner:1995:heliumSpec,CancioPastor:2012:PRL108,Rooij:2011:HeSpectroscopy,Zheng:2017:He_Iso}. The value of the isotope shift from the CREMA collaboration will therefore test the discrepancy with a completely independent method.\\

\textit{Note added in proof}: After submission of this article, Korzinin \etal\ \cite{Korzinin:2018:lamb} was published. A comparison of the radius-independent Lamb shift total in Ref.\,\cite{Korzinin:2018:lamb} (Tab.\,IX: 1668.51(2)\mev\footnote{This value is the radius-independent part of the sum in Tab.\,IX plus two not listed values from Tab.\,VII.}) with our value (Eq.\,\ref{eq:LS:tot}: 1668.489(14)\mev) yields good agreement. The radius-dependent sum from Korzinin \etal{} (Tab.\,VIII and IX: $-106.3(5)\rh^2\mev/\fm^2 + 0.16\mev$\footnote{The $\rh^4$ term is added to the constant term using $\rh=1.681\fm$}) also compares well with ours (Eq.\,\ref{eq:finsize:tot}: $-106.354(8)\rh^2\mev/\fm^2 + 0.078(11)\mev$). It is not clear where the large uncertainty of $0.5\mev/\fm^2$ (corresponding to $\sim1.4\mev$) in the radius-dependent part comes from.
The two-photon exchange contributions are not discussed in detail.\\
The fine structure is given by Korzinin \etal{} in Tab.\,X as $146.181(5)\mev$ (using $1.681\fm$ for the $\rh^2$ term), which agrees with our value (Eq.\,\ref{eq:fs:tot}: $146.1828(3)\mev$). The inconsistency in the Dirac term (\#f1-\#f4) between the Martynenko group and Borie (see Tab.\,\ref{tab:theory0}) is resolved by Korzinin \etal{} which agree with the Martynenko group.\\
\\

\textit{Acknowledgments}: We thank 
E.~Borie, 
A.~P.~Martynenko, 
S.~Karshenboim, 
S.~Bacca,
N.~Barnea, 
M.~C.~Birse,
C.~E.~Carlson,
M.~I.~Eides,
J.~Friar,
M.~Gorchtein,
F.~Hagelstein,
O.~J.~Hernandez,
U.~Jentschura,
C.~Ji,
J.~A.~McGovern,
N.~Nevo Dinur, 
C.~Pachucki,
V.~Pascalutsa,
and M.~Vanderhaeghen 
for fruitful discussions in the creation of this summary.

We specially thank A.~P.~Martynenko for providing the tools to repeat their
calculation of the Friar moment, 
I.~Sick for providing us with a sum of Gaussians FF parameterization of the world data from elastic electron scattering, 
and S.~Bacca, N.~Barnea, O.~J.~Hernandez, C.~Ji, and N.~Nevo Dinur for enlightening discussions regarding the TPE contribution and for making available to us their updated results ahead of publication.

The authors acknowledge support from the European Research Council
(ERC) through StG. \#279765 and CoG. \#725039, the Excellence Cluster PRISMA of the University of Mainz, and the Swiss National Science Foundation SNF, Project 200021\_165854.

\clearpage

\bibliographystyle{mysty1}
\bibliography{ref}

\clearpage
\begin{turnpage}
  \begin{table}[t]
    \begin{center}
      \scalebox{1}{
        \begin{threeparttable}
\setlength\extrarowheight{2.pt}
\centering
\fontsize{5}{6}\selectfont
\begin{tabular}{|ll f{6} l f{7} l f{7} l f{6} l  f{3} l l l|}
\lft{12}{nuclear structure independent terms}\\
\hline
\#{} & Contribution                                       & \cntr{2}{Martynenko group (M)}
                                                                                  & \cntr{2}{Borie (B)}                  
                                                                                                         & \cntr{2}{Karshenboim group (K)} 
                                                                                                                                 & \cntr{2}{Jentschura group (J)}
                                                                                                                                                         & \cntr{2}{Our Choice}          & & Ref.\\
     &                                                    & \cntr{2}{Krutov \etal{} \cite{Krutov:2014:JETP120_73}}
                                                                                  & \cntr{2}{Borie \cite{Borie:2014:arxiv_v7}}                  
                                                                                                         & \cntr{2}{Karshenboim \etal{} \cite{Karshenboim:2012:PRA85_032509,Karshenboim:2010:JETP_LBL}} 
                                                                                                                                 & \cntr{2}{Jentschura \etal{} \cite{Jentschura:2011:SemiAnalytic,Jentschura:2011:PRA84_012505}}    & & & & \\
     &                                                    & &
                                                                                  & &            
                                                                                                         & \cntr{2}{Korzinin \etal{} \cite{Korzinin:2013:PRD88_125019}} 
                                                                                                                                 & & & & & & \\
\hline
\hline
 1 & Non-Rel.\ one-loop electron VP (eVP)                  & 1665.7730             & \#1                  &                       &                       & 1665.7729    & \cite{Karshenboim:2012:PRA85_032509} Tab.\,I           & 1665.772    & \cite{Jentschura:2011:PRA84_012505} Sec.\,II          &                                & & & \\
 2 & Rel.\ corr.\ to one-loop (Breit-Pauli)                 &    0.5210             & \#7, 10              &                       &                       & 0.52110      & \cite{Karshenboim:2012:PRA85_032509} Tab.\,IV; \#4     & 0.521104    & \cite{Jentschura:2011:SemiAnalytic} Eq.\,17d          &                                & & & \\
 3 & Rel.\ one-loop eVP                                    &                       &                      & 1666.305              & Tab.\,5; \#1           &                                             &             &                              & &                              & & & \\   
19 & Rel.\ RC to eVP \app{}\Zap{}$^4$                      &                       &                      &   -0.0090             & Tab.\,5; \#11          &                                             & &           &                              &                                & & & \\    
   & Sum: One-loop eVP (total)\tnote{a}                   & 1666.2940             &                      & 1666.296              &                       & 1666.2940    & \cite{Korzinin:2013:PRD88_125019} Tab.\,I; \#1      & 1666.293    &                              & 1666.2946  & $\pm$0.0014         & & AVG Eq.\,(\ref{eq:LS:leading})\\
\hline
\hline
 4 & Two-loop eVP  (K\"allen-Sabry)                       &   11.5693             & \#2                  &   11.573              & Tab.\,5; \#2           &              &                              &             &                              &                                & & & \\  
 5 & One-loop eVP in 2~C-lines \app{}$^2$\Zap{}$^5$       &    1.7075             & \#9                  &  1.709\tnote{b}     & Tab.\,5; \#3           &              &                              & 1.707588    & \cite{Jentschura:2011:SemiAnalytic} Eq.\,13d          &                                & & & \\
   & Sum: Two-loop eVP \tnote{c}                          &   13.2768             &                      &   13.282              &                       & 13.2769      & \cite{Korzinin:2013:PRD88_125019} Tab.\,I; \#2      &             &                              & 13.2794    & $\pm$0.0026         &  & AVG Eq.\,(\ref{eq:avg:45}) \\
\hline
\hline
6+7 & Third-order eVP                                     &   0.0703              & \#4, 11, 12          &  0.074(3)\tnote{b}  & Tab.\,5; \#4           & 0.074(3)     & \cite{Korzinin:2013:PRD88_125019} Tab.\,I; \#3      &             &                              & 0.0740     & $\pm$0.0030           & & K Eq.\,(\ref{eq:avg:67})\\ 
\hline
\hline
 29 & Second-order eVP contrib.\ \app{}$^2$\Zap{}$^4$m     &  0.0021               & \#8, 13              &                       &                       &  0.00572     & \cite{Korzinin:2013:PRD88_125019} Tab.\,VIII eVP2   &             &                              & 0.0039     & $\pm$0.0018           & & AVG \\        
\hline 
\hline              
 9 & 1:3 LbL (Wichmann-Kroll)                             &  (-0.0199) \tnote{d}  & \#5;                 & -0.01984              &                       & -0.01995(6)  & \cite{Karshenboim:2010:JETP_LBL} Tab.\,III; \#1   &             &                              &                               & &  & \\     
10 & 2:2 LbL (Virtual Delbr\"uck)                         &                       &                      &                       &                       &  0.0114(4)   & \cite{Karshenboim:2010:JETP_LBL} Tab.\,III; \#2   &             &                              &                               & &  & \\     
9a & 3:1 LbL                                              &                       &                      &                       &                       & -0.0050(2)   & \cite{Karshenboim:2010:JETP_LBL} Tab.\,III; \#3   &             &                              &                               & &  & \\   
   & Sum: Light-by-light scattering\tnote{e}              &  -0.0135\tnote{f}   &                      &  -0.0136\tnote{e}   &                       & -0.0136(6)   &                              &             &                              & -0.0136    & $\pm$0.0006        & & K\\      
\hline
\hline
20 & $\mu{}$SE and $\mu{}$VP                              &  -11.1070             & \#24                 &   -11.105708          & Tab.\,2                &              &                              &             &                              & -11.1064   & $\pm$0.0006        & & AVG\\  
11 & $\mu{}$SE corr.\ to eVP \app{}$^2$\Zap{}$^4$           &  -0.0646\tnote{b}   & \#28 Eq.\,99          &  -0.1314\tnote{g}     & App.\,C Tab.\,16        &  -0.06462    & \cite{Korzinin:2013:PRD88_125019} Tab.\,VIII; a     & -0.06462    & \cite{Jentschura:2011:SemiAnalytic} Eq.\,29d          & -0.0646                       & & & K, J\\  
\hline
\hline   
12 & eVP loop in SE \app{}$^2$\Zap{}$^4$                  &  -0.0307              & \#27                 &                       &                       &  -0.03073    & \cite{Korzinin:2013:PRD88_125019} Tab.\,VIII; d     &             &                              &                               & & & \\
30 & Hadr.\ loop in SE \app{}$^2$\Zap{}$^4$m               &                       &                      &                       &                       &  -0.00041(4) & \cite{Korzinin:2013:PRD88_125019} Tab.\,VIII; e     &             &                              &                               & & & \\        
13 & Mixed eVP + $\mu{}$VP                                &  0.0023               & \#3                  &   0.00208             & Tab.\,5                &   0.00395    & \cite{Korzinin:2013:PRD88_125019} Tab.\,VIII; b     &             &                              &                               & & & \\
31 & Mixed eVP + had.\ VP                                   &                       &                      &                       &                       &   0.0025(2)  & \cite{Korzinin:2013:PRD88_125019} Tab.\,VIII; c     &             &                              &                               & & & \\
21 & Higher ord.\ corr.\ to $\mu{}$SE/$\mu{}$VP             &                       &                      &  -0.034663            & Tab.\,2                &              &                              &             &                              &                               & & &\\
   & Sum: Higher orders \tnote{h}                         & -0.0284               &                      &  -0.032583            &                       & -0.02469(20) &                              &             &                              & -0.0286    & $\pm$0.0039        & & AVG\\
\hline
\hline  
14 & Hadronic VP                                          &  0.2229               & \#29                 & 0.228(12)             & Tab.\,5; \#7           &              &                              &             &                              & 0.2280     & $\pm$0.0120        & & B\\                      
\hline
\hline   
17 & Recoil corr.\ \Zap{}$^4$m$^3$/M$^2$ (Barker-Glover)   &  0.2952               & \#21                 &  0.2952               & Tab.\,5; \#12          &              &                              &  0.29518    & \cite{Jentschura:2011:SemiAnalytic} Eq.\,A3           & 0.2952                        & & & B, M, J\\
18 & Recoil finite size\tnote{i}                          &                       &                      &  0.2662(1)            & Tab.\,5; \#10          &              &                              &             &                              &                               & & & \\
22 & Rel.\ RC \Zap{}$^5$                                   & -0.4330               & \#22                 & -0.4330               & Tab.\,5; \#8           &              &                              &  -0.433032  & \cite{Jentschura:2011:SemiAnalytic} Eq.\,32d          & -0.4330                       & & & B, M, J\\
23 & Rel.\ RC \Zap{}$^6$                                   & 0.0038                & \#23                 &                       &                       &              &                              &             &                              & 0.0038                        & & & M\\
24 & Higher ord.\ rad.\ recoil corr.                        & -0.0377\tnote{f}      & \#25                 & -0.04737              & Tab.\,5; \#9; p.9      &              &                              &             &                              & -0.0474                       & & & B \\    
28 & Rad.\ (only eVP) RC \app{}\Zap{}$^5$                  &                       &                      &                       &                       &              &                              &   0.003867  & \cite{Jentschura:2011:SemiAnalytic} Eq.\,46d          & 0.0039                        & & & J\\
\hline
\hline 
   & \bf{Sum}                                             &  1668.4809             &                     & 1668.8641(50)\tnote{j}&                       &              &                              &             &                              & 1668.4892  & $\pm$0.0135        & & Eq.\,(\ref{eq:LS:tot}) \\
\hline
\end{tabular}
\par\medskip
\begin{tablenotes}
\item[a] sum: 1+2 or 3+19
\item[b] taken from \cite{Karshenboim:2012:PRA85_032509}
\item[c] sum: 4+5
\item[d] not part of final sum
\item[e] sum: 9+9a+10
\item[f] taken from \cite{Karshenboim:2010:JETP_LBL}
\item[g] incomplete
\item[h] sum: 12-21
\item[i] included in elastic TPE
\item[j] taken from text in~\cite{Borie:2014:arxiv_v7} p.\,14
\end{tablenotes}
\end{threeparttable}

      }
      \caption{
        Terms contributing to the \MuFourHe{} Lamb shift $\Delta E(\mathrm{2P_{1/2}-2S_{1/2}})$ that are independent of the nuclear structure. The last column shows the values used for our Lamb shift determination. Uncertainties for the different terms are either given in the respective publication or represent the spread of the different calculations. All values are given in meV. 
      }
      \label{tab:theory1}
    \end{center}
  \end{table}
\end{turnpage}
\clearpage
\begin{turnpage}
  \begin{table}[t]
    \begin{center}
      \scalebox{1}{
        \begin{threeparttable}
\setlength\extrarowheight{2.pt}
\centering
\fontsize{5}{6}\selectfont
\begin{tabular}{|ll f{7} l f{9} l f{4} l f{4} llll|}
\lft{12}{nuclear structure dependent terms}\\
\hline
\#{} & Contribution                                              & \cntr{2}{Martynenko group (M)}     & \cntr{2}{Borie (B)}          & \cntr{2}{Karshenboim group (K)} &  \cntr{2}{Our Choice}              & & & Ref.  \\
     &                                           & \cntr{2}{Krutov \etal{} \cite{Krutov:2014:JETP120_73}}     & \cntr{2}{Borie \cite{Borie:2014:arxiv_v7}}          & \cntr{2}{Karshenboim \etal{} \cite{Karshenboim:2012:PRA85_032509}} &  &  & & & \\
\hline
\hline
 r1 & Non-Rel.\ finite size \Zap{}$^4$                            &  -105.323\hphantom{0}\rs & \#14; Eq.\,61                                   & -105.319\hphantom{0}\rs  & Tab.\,14b$_a$          &  -105.32~\,\rs{} &~~Tab.\,III;  \#1    &       -105.3210       & $\pm$0.0020 &\rs        & & AVG Eq.\,(\ref{eq:finsize:r1}) \\
\hline
\hline
 r4 & Uehling corr.\ (+KS), \app{}\Zap{}$^4$               &  -0.3414\rs              & \#16; Eq.\,69                                   &                    &                                 & -0.333\rs        &~~Tab.\,III; \#3     &                      &         &        & & \\        
 r6 & Two-loop eVP corr.\ \app{}$^2$\Zap{}$^4$                     &  -0.0027\rs              & \#18; Eq.\,73                                   &                    &                                &                  &                    &                            &             &        & &  \\        
sum & r4+r6                                                       &  -0.3441\rs              &                                         &  -0.3297\rs         &  Tab.\,14b$_d$                              &                  &                    &       -0.3369         & $\pm$0.0072   &\rs        & & AVG \\        
\hline
\hline
 r5 & One-loop eVP in SOPT, \app{}\Zap{}$^4$       &  -0.5362\rs              & \#17; Eq.\,70                                   &                  &                 & -0.536\rs        &~~Tab.\,III; \#2     &                        &                       & & \\    
 r7 & Two-loop eVP in SOPT, \app{}$^2$\Zap{}$^4$                 &  -0.0065\rs              & \#19                                 &                    &                                &                  &                    &                   &             &         & & \\ 
sum & r5+r7                                                    &  -0.5427\rs              &                                         &  -0.5392\rs        & Tab.\,14b$_e$                         &                  &                    &       -0.5410         &  $\pm$0.0017          &\rs        & & AVG\\ 
\hline
\hline
 r8 & Corr.\ to 2P$_{1/2}$ level                                   &                          &                                          & +0.004206\rs~\tnote{a}   & Tab.\,14b(2p1/2)       &                  &                    &       0.0042          &             &\rs        & & B\\   
 r2 & Rad.\ corr.\ finite size \app{}\Zap{}$^5$                    &  -0.0250\rs              & \#26; Eq.\,92                          &   -0.0250\rs       & Tab.\,14b$_b$                &                  &                    &       -0.0250        &             &\rs        & & B, M\\
 r3 & Finite size corr.\ \Zap{}$^6$                               &  -0.1370\rs              & \#26; Eq.\,91(1)                        &   -0.1310\rs       & Tab.\,14b$_c$                &                  &                    &       -0.1340        & $\pm$0.0030 &\rs        & & AVG \\
\hline
\hline 
    & Sum of coefficients                                         & -106.372\hphantom{0}\rs &                                          & -106.340\hphantom{0}\rs   &                          &                  &                    &    -106.3536(82)      &             &  \rs      & & \\   
\hline
\hline
r2' & Rad.\ corr.\ $\alpha$\Zap{}$^5$                            &  0.0112                 & \cite{Faustov:2017:rad_fin_size}         &                    &                                 &                  &                    &   0.0112          &             &           & & M\\
r2b'& VP corr.\ $\alpha(Z\alpha)^5$~\tnote{b}              &  0.1270(13)        & \#20; Eq.\,78                        &              &                 &                  &                    &                        &            &           & & \\
r3' & Remaining \Zap{}$^6$ contrib.                              &  0.07846                & \#26; Eq.\,91(2)                        &  0.056             & Tab.\,5; \#17                 &                  &                    &       0.0672           & $\pm$0.0112  &           & & AVG\\
\hline
\hline 
    & \textbf{Sum}                                               & \multicolumn{2}{l}{~-106.3718\rs+0.2167}                            & \multicolumn{2}{l}{~~-106.3397\rs+0.056}              &                  &                    & \multicolumn{5}{l|}{~~-106.3536(82)\rs+0.0784(112)~~Eq.\,(\ref{eq:finsize:tot})}  \\   
\hline
\end{tabular} 
\par\medskip
\begin{tablenotes}
\item[a] sign changed due to convention
\item[b] not included in our choice, see text.

\end{tablenotes}
\end{threeparttable}

      }
      \caption{Nuclear structure dependent terms to the \MuFourHe{} Lamb shift $\Delta E(\mathrm{2P_{1/2}-2S_{1/2}})$ with their respective charge radius scaling.  The values are given in meV/fm$^2$, except \#r2', \#r2b', and \#r3', which are given in \mev.
                }
                \label{tab:theory2}
    \end{center}
  \end{table}
\end{turnpage}
\clearpage
\begin{turnpage}
  \centering
  \begin{table}[t]
    \begin{center}
      \scalebox{1}{
        \begin{threeparttable}
\setlength\extrarowheight{2.pt}
\centering
\fontsize{5}{6}\selectfont
\begin{tabular}{|ll f{5} l f{5} l f{5} l f{3} lll|}
\lft{10}{fine structure}\\
\hline
\#{} & Contribution                                        & \cntr{2}{Martynenko group (M)}        & \cntr{2}{Borie (B)}         & \cntr{2}{Karshenboim group (K)} & \cntr{2}{Our Choice}              & & Ref.\\
     &                                       & \cntr{2}{Elekina \etal{} \cite{Elekina:2011:mu4He}} & \cntr{2}{Borie \cite{Borie:2014:arxiv_v7}} & \cntr{2}{Karshenboim \etal{} \cite{Karshenboim:2012:PRA85_032509}} & & & & \\
     &                                       & & & & & \cntr{2}{Korzinin \etal{} \cite{Korzinin:2013:PRD88_125019}} &  & & & \\
\hline
\hline    
 f1 & Dirac                                                &                                    & & 145.7183         & Tab.\,7; \#1       &                                     & &                                 & & & \\    
 f2 & Recoil corr.                                         &                                    & &  -0.1107         & Tab.\,7; \#6       &                                     & &                                 & & & \\  
 f3 & \Zap{}$^4$ contrib.                                  &  145.56382  & \#1                    &                                    & &                                     & &                                 & & & \\
 f4a & \Zap{}$^6$ contrib.                                 &  0.01994    & \#3                    &                                    & &                                     & &                                 & & & \\
 f4b  & \Zap{}$^6$m$_1$/m$_2$ contrib.                     &  -0.00045   & \#4                    &                                    & &                                     & &                                 & & & \\
   & Sum: Dirac \tnote{a}                                  &  145.58331                         & & 145.6076                           & &                                     & &  145.5833 &                 & & M\\
\hline
\hline    
 f5a & One-loop eVP (Uehling), \app{}\Zap{}$^4$            &  0.13167   & \#5                     &                                    & &                                     & &                                 & & & \\   
 f5b & One-loop eVP in FS interaction                      &  0.14396   & \#7                     &                                    & &                                     & &                                 & & & \\
     & Sum: One-loop eVP \tnote{b}                         &  0.27563                           & &   0.2753         & Tab.\,7; \#2       & 0.27502 & \cite{Karshenboim:2012:PRA85_032509} Tab.\,IV         & 0.2753 & $\pm$0.0003          & & AVG\\
f6a & Two-loop eVP (K\"allen-Sabry)                        &  0.00097   & \#10,11                 &                  &      &                                     &                        & 0.0010     & & & M\\
f6b & Two-loop eVP in 2 C-lines                            &  0.00231   & \#9,12,13               &  0.0021        & Tab.\,7; \#3  &  0.00247 & \cite{Korzinin:2013:PRD88_125019} Tab.\,VII        &   0.0025   & & & K\\
f7 & \app{}$^2$\Zap{}$^4$m contrib.                        &                                    & &                                    & &  0.0023 & \cite{Korzinin:2013:PRD88_125019} Tab.\,IX(a,d,e)        & 0.0023 &                         & & K \\
 f11* & \app{}\Zap{}$^6$ contrib.                          & -0.00056    & \#8                    &                                    & &                                     & &     -0.0006                            & & & M\\       
f12* & One-loop $\mu{}$VP                                  &  0.00001   & \#6                     &                                    & &                                     & &  0.0000                        & & & M\\
\hline
\hline 
f8 & $\mu{}$ anom.\ mag.\ mom.\ ($a_{\mu{}}$)                &                                    & & 0.3290           & Tab.\,7; \#4       &                                     & &                                 & & & \\ 
f9 & $\mu{}$ anom.\ mag.\ mom.\ higher orders                 &                                    & & 0.0013           & Tab.\,7; \#5       &                                     & &                                 & & & \\   
   & Sum: $\mu{}$ anom.\ mag.\ mom.\ \tnote{c}                &  0.33032   & \#2                     & 0.3303                             & &                                     & & 0.3303                         & & & B, M\\     
\hline
\hline   
f10a & Finite size (1st ord.)                              &  -0.01176  & \#14                    & -0.0119          & Tab.\,7; \#7       &                                     & & -0.0118 & $\pm$0.0001           & & AVG \\    
f10b & Finite size (2nd ord.)                              &   0.00045  & \#15                    &                                    & &                                     & &  0.0005                        & & & M  \\      
\hline
\hline   
  & \bf{Sum}                                               & 146.18068  &                         &  146.2034        &                  &                                     & & 146.1828 & $\pm$0.0003          & & Eq.\,(\ref{eq:fs:tot})\\
\hline
\end{tabular}
\par\medskip
\begin{tablenotes}
\item[a] sum: f1+f2 / f3+f4a+f4b
\item[b] sum: f5a+f5b
\item[c] sum: f8+f9
\item[*] new term
\end{tablenotes}
\end{threeparttable}

      }
      \caption{ Terms contributing to the fine structure $\Delta E(\mathrm{2P_{3/2}-2P_{1/2}})$ in \MuFourHe{} taken from different sources. All values are given in meV.}
                \label{tab:theory0}
    \end{center}
  \end{table}
\end{turnpage}
\clearpage

\end{document}